\documentclass[11pt]{article}
\usepackage{vmargin,graphicx,amsmath,calc,epsf,epsfig,subfigure,array,rotating,rotating,rotate,tabularx,pifont,multirow}
\usepackage{indentfirst}
\usepackage{rotate}
\usepackage{float}

\newcommand{\BABARPubYear}    {03}

\newcommand{\BABARConfNumber} {08}
\newcommand{\SLACPubNumber} {9696}

\input babarsym
\def\sintwob {\ensuremath{\sin{2\beta}}}
\def\Dstarp  {\ensuremath{D^{*+}}}
\def\Dstarm  {\ensuremath{D^{*-}}}

\def\Dstarz  {\ensuremath{D^{*0}}}
\def\Dp      {\ensuremath{D^+}}
\def\Dm      {\ensuremath{D^-}}

\def\Dz      {\ensuremath{D^0}}

\newcommand{\re}{{$\mathop{\rm Re}$}}
\newcommand{\taub}{\ensuremath{\tau_B}}

\newcommand{\dGoverG}{\ensuremath{ \Delta \Gamma/\Gamma }}
\newcommand{\dGoverGd}{\ensuremath{ \Delta \Gamma/\Gamma }}
\newcommand{\sgndGoverG}{\ensuremath{ \sign(\relambcpbare) \Delta \Gamma/\Gamma }}
\newcommand{\sgndGoverGd}{\ensuremath{ \sign(\relambcpbare) \Delta \Gamma/\Gamma }}
\newcommand{\dG}{\ensuremath{ \Delta \Gamma }}
\newcommand{\dGd}{\ensuremath{ \Delta \Gamma }}

\newcommand{\dM}{\ensuremath{ \Delta m }}

\newcommand{\absqop}{\ensuremath{|q/p|}}

\def\lambtagbare {\ensuremath{ \lambda_{\rm tag} }}

\def\lambflavbare {\ensuremath{ \lambda_{\rm flav} }}
\def\lambbarflavbare {\ensuremath{ \overline{\lambda}_{\rm flav} }}

\def\imlambflavbare {\ensuremath{ \im \lambda_{\rm flav} }}
\def\imlambbarflavbare {\ensuremath{ \im \overline{\lambda}_{\rm flav} }}
\def\relambtagbare {\ensuremath{ \re \lambda_{\rm tag} }}
\def\relambbartagbare {\ensuremath{ \re \overline{\lambda}_{\rm tag} }}
\def\relambflavbare {\ensuremath{ \re \lambda_{\rm flav} }}
\def\relambbarflavbare {\ensuremath{ \re \overline{\lambda}_{\rm flav} }}

\def\lambflavbarbare {\ensuremath{ \lambda_{\overline{\rm flav}} }}

\def\imlambtagflat {\ensuremath{ \im \lambda_{\rm tag} /|\lambda_{\rm tag}| }}

\def\imlambbartagflat {\ensuremath{ \im \overline{\lambda}_{\rm tag} /|\overline{\lambda}_{\rm tag}| }}

\def\imlambflavflat {\ensuremath{ \im \lambda_{\rm flav} /|\lambda_{\rm flav}| }}

\def\imlambbarflavflat {\ensuremath{ \im \overline{\lambda}_{\rm flav} /|\overline{\lambda}_{\rm flav}|}}

\def\imlambcpflat {\ensuremath{ \im \lambda_{\CP} /|\lambda_{\CP}|}}

\def\lambcpbare {\ensuremath{ \lambda_{\CP} }}
\def\imlambcpbare {\ensuremath{ \im \lambda_{\CP}  }}
\def\relambcpbare {\ensuremath{ \re \lambda_{\CP}  }}

\newcommand{\ckm}{\ensuremath{ \rm CKM }}
\newcommand{\dckm}{\ensuremath{ \rm DCKM }}

\newcommand{\imZ}{\ensuremath{\im z}}
\newcommand{\reZ}{\ensuremath{\re z}}

\newcommand{\reZparflat}{\ensuremath{\left( \relambcpbare / |\lambcpbare| \right) \reZ }}

\def\dw   {\ensuremath{\Delta w}}

\def\dwa   {\ensuremath{\Delta w^\alpha}}
\def\wa   {\ensuremath{w^\alpha}}

\def\waj   {\ensuremath{w^{\alpha,j}}}

\def\bflav   {\ensuremath{B_{\rm flav}}}
\def\bcpks   {\ensuremath{B_{{\CP}\KS}}}
\def\bcpkl   {\ensuremath{B_{{\CP}\KL}}}
\def\bcp   {\ensuremath{B_{\CP}}}
\def\btag   {\ensuremath{B_{\rm tag}}}
\def\brec   {\ensuremath{B_{\rm rec}}}

\def\mes   {\ensuremath{m_{\rm ES}}}
\def\de   {\ensuremath{\Delta E}}

\def\bflavz   {\ensuremath{B_{\rm flav}^0}}
\def\bflavzb  {\ensuremath{\overline{B}_{\rm flav}^0}}

\def\flavz   {\ensuremath{\rm flav}}
\def\flavzb   {\ensuremath{\rm \overline{flav}}}
\def\cpks   {\ensuremath{{\CP}}\KS}
\def\cpkl   {\ensuremath{{\CP}}\KL}
\def\cp   {\ensuremath{\CP}}
\def\tagz   {\ensuremath{\rm tag}}
\def\tagzb   {\ensuremath{\rm \overline{tag}}}

\def\sign{\mathop{\rm{sgn}}} 
\def\Rcp {\ensuremath{ R_{\CP} }}

\newif\ifdimspec
\def\figsize#1{\dimspecfalse \checkdim#1\end
\ifdimspec
  \def\figureWidth{#1}%
\else
  \def\figureWidth{#1 in}\fi}
\def\checkdim#1{\ifx#1\end \let\next=\relax
  \else \ifcat#1a \dimspectrue \fi \let\next=\checkdim\fi \next}
\newcommand{\lblcaption  }[2]{\caption{#2\label{\secname#1}}}
\newcommand{\twoAngFiguresEPS}[6]
{
\begin{figure}
\begin{center}
\figsize{#4}
\begin{minipage}[t]{3.2in}
\begin{center}
\epsfig{file=\sectiondir/#1.eps,width=\figureWidth,angle=#5}
\end{center}
\end{minipage}
\begin{minipage}[t]{3.2in}
\begin{center}
\epsfig{file=\sectiondir/#2.eps,width=\figureWidth,angle=#6}
\end{minipage}
\end{center}
\lblcaption{#1}{#3}
\end{figure}
}
\newcommand{\pvec}{{\bf p}} 

\newcommand{\ket}[1]{\ensuremath{|{#1}\rangle}}
  
\def\beq{\begin{equation}}
\def\eeq{\end{equation}}
\def\bea{\begin{eqnarray}}
\def\eea{\end{eqnarray}}
\def\bq{\begin{quote}}
\def\eq{\end{quote}}
\def\ben{\begin{enumerate}}
\def\een{\end{enumerate}}
\def\nn{\nonumber}

\def\re{{\mathop{\rm Re}}}
\def\im{{\mathop{\rm Im}}}

\def\kl{$\KL$}
\def\dg{\Delta \Gamma}

\def\dt{\Delta t}

\def\sdt{\sigma_{\dt}}

\def\Bzd    {\ensuremath{B^0_d}}

\def\dt   {\ensuremath{\Delta t}}
\def\dz   {\ensuremath{\Delta z}}

\def\Dstarp   {\ensuremath{D^{*+}}\xspace}
\def\Dstarm   {\ensuremath{D^{*-}}\xspace}

\long\def\inst#1{\par\nobreak\kern 4pt\nobreak
    {\it #1}\par\vskip 10pt plus 3pt minus 3pt}

\begin{document}
{\pagestyle{empty}

\begin{flushright}
\hfill
BABAR-CONF-\BABARPubYear/\BABARConfNumber \\
SLAC-PUB-\SLACPubNumber \\
March 2003 \\
\end{flushright}

\par\vskip 0.5cm

\begin{center}
\boldmath
\Large \bf Limits on the Lifetime Difference of Neutral \B\ Mesons 
and on \CP, \T, and \CPT Violation in \BzBzb Mixing 
\end{center}
\bigskip

\begin{center}
\large The \babar\ Collaboration\\
\mbox{ }\\
\today
\end{center}
\bigskip \bigskip

\begin{center}
\large \bf Abstract
\end{center}

Using events in which one of two neutral \B\ mesons from the decay of
an \FourS\ resonance is fully reconstructed, we set limits on
the lifetime difference between the two neutral-\B\ mass eigenstates
and on \CP, \T, and \CPT\ violation in $\BzBzb$ mixing.  
Both \CP\ and non-\CP\ eigenstates were obtained from the 88 million
$\FourS \to \B\Bb$ decays collected between 1999 and
2002 with the \babar\ detector at 
the \pep2\ asymmetric-energy \B-Factory at SLAC.  We determine
six independent parameters governing mixing (\dM, \dGoverG), 
\CPT/\CP\ violation \mbox{(\reZ, \imZ)}, and \CP/\T\ violation
(\imlambcpbare, \absqop), where $\lambcpbare$ characterizes \Bz\ and \Bzb\ decays to
 states of charmonium plus \KS\ or \KL. The preliminary results are
$$\begin{array}{
r@{\ \ =\ }r@{.}l@{\pm 0.}l@{{\rm (stat.)}\pm 0.}
l@{{\rm{(syst.)}}\ \ [}r@{,}l@{]}l}
\sgndGoverGd&-0&008&037&018&-0.084&0.068&~,\\
\absqop     &1 &029&013&011& 1.001&1.057&~,\\
\reZparflat &0 &014&035&034&-0.072&0.101&~,\\
\imZ        &0 &038&029&025&-0.028&0.104&~.
\end{array}$$
The values inside square brackets indicate the 90\% confidence-level
intervals. For \mbox{\imlambcpflat} and \dM\ we find values consistent with recent
results from other analyses. These results are consistent with
Standard Model expectations.

\vfill
\begin{center}
Presented at the XXXVIII$^{th}$ Rencontres de Moriond on\\
Electroweak Interactions and Unified Theories, \\
3/15--3/22/2003, Les Arcs, Savoie, France
\end{center}

\begin{center}
{\em Stanford Linear Accelerator Center, Stanford University, 
Stanford, CA 94309} \\ \vspace{0.1cm}\hrule\vspace{0.1cm}
Work supported in part by Department of Energy contract DE-AC03-76SF00515.
\end{center}

\newpage
} 

\begin{center}
\small

The \babar\ Collaboration,
\bigskip

%
B.~Aubert,
R.~Barate,
D.~Boutigny,
J.-M.~Gaillard,
A.~Hicheur,
Y.~Karyotakis,
J.~P.~Lees,
P.~Robbe,
V.~Tisserand,
A.~Zghiche
\inst{Laboratoire de Physique des Particules, F-74941 Annecy-le-Vieux, France }
% ...

\end{center}\newpage


\section{\boldmath Introduction and analysis overview}
\label{sec:intro}

The neutral \Bzd\ meson system has two mass eigenstates with mass and
total decay rate differences \dM\ and \dGd. While the mass difference
has been measured recently with high precision
\cite{ref:dM-babar-had,ref:dM-babar-dstlnu,ref:dM-babar-dilep,ref:dM-belle}, 
only weak limits exist for the lifetime difference
$\Delta\taub=-\Delta\Gamma/\Gamma^2$. Using the time-integrated mixing
parameter $\chi_d$, the CLEO Collaboration has set a limit of 
$|\Delta \Gamma/\Gamma | <80\%$ 
\cite{ref:cleochid}.  A stronger constraint, 
$| \dGoverGd | <20\%$ at 90\% confidence-level,
has been obtained by the DELPHI Collaboration from a direct
time-dependent study using flavor eigenstate events \cite{ref:DELPHI}.
In the Standard Model, the ratio of the difference in the decay widths
to the difference of the masses is proportional to $m_b^2/m_t^2$ and
thus quite small.
Recent calculations of \dGoverGd, including $1/m_b$ contributions and
part of the next-to-leading order QCD corrections within the Standard
Model
\cite{ref:dighe}, find values of about $-0.3\%$.
The large data set available at the asymmetric \B\ Factories provides
the opportunity to reach much closer to the anticipated range for
$\Delta \Gamma/\Gamma$.

The behavior of neutral \B\ mesons is 
sensitive to \CPT\ violation~\cite{ref:ko,ref:bbCPT}.  The
\CPT\ theorem \cite{ref:cpttheo,ref:cpttheo2}, based on very general
principles of relativistic quantum field theory, states that 
the triple product of the universal discrete symmetries \C, \P, 
and \T\ represent an exact symmetry. The \CPT\ symmetry remains to date
the only combination of \C, \P, and  \T\ that is not known to be violated.
However, the proof of the \CPT\ theorem relies on locality, which could
break down at very short distances. For instance, string theories are
fundamentally non-local and therefore do not necessarily fulfill the
conditions of the \CPT\ theorem. Therefore it is possible, although
perhaps unlikely, that \CPT\ could break down.  To date, the best tests have
come from experiments in the neutral kaon system \cite{ref:cptkaons}. 
Bounds obtained so far in the \B\ meson system 
\cite{ref:dM-belle,ref:otherCPTBtests} are,
however, mainly sensitive to the absorptive (lifetime) component of the 
Hamiltonian, where the small expected value of \dGd\ suppresses the asymmetry effects.

Violation of \CP\ in the neutral \B\ meson system may occur in mixing,
in decay, or in the interference between mixing and decay.
There is no fundamental way of assigning the source of \CP\ violation
observed in interference to either mixing or to decay. The standard phase
choice puts \CP\ violation in the mixing, but this is simply a convention.
Other observable processes, however, can isolate \CP\ violation due
entirely to mixing.  Similarly, mixing may intrinsically
contain \T\ violation or even \CPT\ violation.  It is these possibilities
for the breaking of discrete symmetries in mixing itself that we
address in this analysis.

To measure the lifetime difference of the neutral \B-meson mass
eigenstates and  \CP, \T, or \CPT\ violation we observe the time
dependence of decays of neutral \B\ mesons produced in pairs at
the \FourS\ resonance.  The conventional mixing and \CP\ analyses allow for
exponential decay modulated by oscillatory terms with frequency
\dM.
This neglects the difference between the decay rates
\dG\ of the two mass eigenstates, which would introduce new
exponential factors.  \CP, \T, and \CPT\ violation in the mixing of the
neutral \B\ mesons would modify the coefficients of the various terms
involving exponential and oscillatory behavior.  To detect these
potential subtle changes requires precision measurements of the decays
and thorough consideration of systematic issues.  It also requires a
more comprehensive treatment of the coherent decays of the mesons than
has been conducted previously.

The analysis is based on a total of about 88 million
$\FourS \to \B\Bb$ decays collected between 1999 and 2002 with the
\babar\ detector at the PEP-II asymmetric-energy
\B\ Factory at the Stanford Linear Accelerator Center. There, 9 \gev\ electrons and 
3.1 \gev\ positrons annihilate to produce the $\B\Bb$ pairs moving
along the $e^-$ beam direction ($z$-axis) with a Lorentz boost of
$\beta \gamma \approx 0.56$, allowing a measurement of the proper time
difference \dt\ between the two \B\ decays.  In this analysis, 
one \B\ meson is fully reconstructed in a flavor (\bflav) or \CP\ (\bcp)
eigenstate (generally denoted as \brec).
The remaining charged particles in the event, which originate from the
other \B\ meson (\btag), are used to identify its flavor as \Bz\ or
\Bzb. The time difference $\dt = t_{\rm rec} - t_{\rm tag} \approx
\Delta z / \beta \gamma c$ is determined from the separation $\Delta
z$ of the decay vertices for the fully reconstructed \B\ candidate and
the tagging \B\ along the boost direction.

A single maximum-likelihood fit to the time distributions of 
tagged and untagged, flavor and \CP\ eigenstates determines
six independent parameters (see Sec.~\ref{sec:decayrates}) governing mixing (\dM, \dGoverGd),
\CPT/\CP\ violation (\reZ, \imZ) and \CP/\T\ violation (\imlambcpbare, \absqop), where 
$\lambcpbare$ is the traditional variable characterizing the decays of neutral 
\B\ mesons into final states of charmonium and a \KS\ or \KL. The parameters
\imlambcpbare\ and \dM\ are
used only as a cross-check with the \babar\ $\sin2\beta$ analysis \cite{ref:sin2b-babar} and
previous \dM\ results \cite{ref:dM-babar-had,ref:dM-babar-dstlnu,ref:dM-babar-dilep,ref:dM-belle}.

The analysis has several challenges.  
First, the tagged \B\ and the fully reconstructed \B\ decays are
correlated and interference between allowed
and doubly-\ckm-suppressed (DCKM) decays cannot be neglected. Second, tagging 
incorrectly assigns the flavor with a certain mistag probability. 
Third, the resolution for \dt\ is comparable to the
\B\ lifetime and asymmetric for positive and negative \dt.
This asymmetry must be well understood lest it be mistaken for a fundamental
asymmetry we seek to measure. 
Fourth, possible direct \CP\ violation in the \bcp\ sample
can be a competing source of fake effects and must be parameterized
appropriately. Finally, we have to account possible asymmetries
induced by the differing response of the detector to positive and
negative particles. In resolving these issues we rely mainly on data.

The paper is organized as follows. In Sec.~\ref{sec:decayrates} we
present a general formulation of the time-dependent decay rates of
$\BzBzb$ pairs produced at the \FourS\ resonance, including
effects from the lifetime difference, possible \CPT\ violation, and
interference effects induced by doubly-\ckm-suppressed decays. 
We derive the expressions for \B\ decays to final states with flavor and \CP\ eigenstates. 
In Sec.~\ref{sec:detector} we
describe the \babar\ detector.  After discussing the data sample 
in Sec.~\ref{sec:sample}, we describe the $b$-quark tagging
algorithm in Sec.~\ref{sec:tagging}. Sec.~\ref{sec:decaytime} is devoted to the description of
the measurement of \deltaz\ and the determination of \dt\ and its
resolution function. In Sec.~\ref{sec:method} we describe the unbinned
log-likelihood function and the assumptions made in the nominal
fit. The results of the fit are given in Sec.~\ref{sec:results}.
Cross-checks are discussed in Sec.~\ref{sec:checks} and systematic
uncertainties are summarized in Sec.~\ref{sec:systematics}. The
results of the analysis are summarized and discussed in
Sec.~\ref{sec:summary}.

\section{\boldmath General time-dependent decay rates from $\Upsilon(4S)\to
\BzBzb$}
\label{sec:decayrates}

The neutral \B\ meson system can be described by the effective Hamiltonian $\tilde{M}-i\tilde{\Gamma}/2$,
where ${\tilde{M}}$ and ${\tilde{\Gamma}}$ are two-by-two hermitian matrices
describing, respectively, the mass (dispersive) and lifetime (absorptive) components.
If either \CP\ or \CPT\ is a good symmetry, then $M_{11}=M_{22}$ and $\Gamma_{11}=
\Gamma_{22}$, with the index $1$ indicating \Bz\  and $2$ indicating
\Bzb. If either \CP\ or \T\ is a good symmetry, $\Gamma_{12}/M_{12}$ is
real.  This condition does not depend on the phase convention 
chosen for the \Bz and \Bzb.  The masses $\mu_\pm$ and decay rates $\gamma_\pm$ of the two eigenstates are
\begin{equation}
\omega_\pm= \mu_\pm -\frac i2\gamma_\pm= M-\frac i2 \Gamma \pm
       \sqrt{\left(M_{12}-\frac i2\Gamma_{12}\right)\left(M_{12}^*-\frac i2\Gamma_{12}^*\right)+\left(\delta M-\frac i2\delta \Gamma\right)^2}~,
\end{equation}
where we define
\begin{equation}
M \equiv \frac{M_{11}+M_{22}}{2} \quad , \quad
\Gamma \equiv \frac{\Gamma_{11}+\Gamma_{22}}{2} \quad , \quad
\delta M \equiv \frac{M_{11}-M_{22}}{2} \quad , \quad
\delta\Gamma \equiv \frac{\Gamma_{11}-\Gamma_{22}}{2}~.
\label{eq:defns}
\end{equation}
Neglecting \CPT\ violation, and anticipating that the lifetime difference is
small compared to the mass difference, we have
\begin{equation}
\dM = 2|M_{12}|~; \qquad \dG = 2|M_{12}| \re  (\Gamma_{12}/M_{12})~.
\end{equation}
Here we have taken $\dM$ to be the mass of the heavier state minus
the mass of the lighter.  Thus $\dG$ is the decay rate of the
heavier state minus the decay rate of the lighter and its sign is not known a priori.

The light and heavy mass eigenstates of the neutral $B$-meson system may
be written
\begin{eqnarray}
|B_L\rangle&=&p|\Bz\rangle +q|\Bzb\rangle\nonumber\\
|B_H\rangle&=&p|\Bz\rangle -q|\Bzb\rangle
\end{eqnarray}
where
\begin{equation}
 {q\over p}\equiv-\sqrt{{M^\ast_{12}-\frac{i}{2}\,\Gamma^\ast_{12}\over 
M_{12}-\frac{i}{2}\,\Gamma_{12}}}~.
\end{equation}
The magnitude of $q/p$  is very nearly unity:
\begin{equation}
\left|\frac qp\right|^2\approx 1-\im \frac{\Gamma_{12}}{M_{12}}~.
\end{equation}
In the Standard Model, the \CP- and \T-violating quantity $|q/p|^2-1$ is
small not just because $|\Gamma_{12}|$ is small, but additionally
because the \CP-violating quantity $\im (\Gamma_{12}/M_{12})$
would vanish if the $u$- and $c$-quark mass were the same.  
\CP\ violation is not possible in mixing if two of the quark masses (for quarks
of identical charge) are
identical because we could redefine them so one quark did not mix with
the other two.  The remaining two generations would be inadequate to
support \CP\ violation.  The result is that $\im(\Gamma_{12}/M_{12})$ is suppressed by an additional factor
$m_c^2/m_b^2\approx 0.1$.  When the remaining factors are included,
the result is $|\im ({\Gamma_{12}}/{M_{12}})|<10^{-3}$.
 
\CPT\ violation in mixing can be described conveniently by the phase-convention independent quantity
\begin{equation}
z\equiv{\delta M-\frac i2\,\delta\Gamma\over
\sqrt{\left(M_{12}-\frac{i}{2}\,\Gamma_{12}\right)
\left(M^\ast_{12}-\frac{i}{2}\,\Gamma^\ast_{12}\right)
+ \left(\delta M - \frac{i}{2}\,\delta\Gamma\right)^2}} ~=~ 
\frac{\delta M-\frac i2\,\delta\Gamma}{\frac12 \left( \dM - \frac i2 \dG \right)}~.\label{eq:zdefn}
\end{equation}
States that begin as purely \Bz\ or \Bzb\  will oscillate and after a time $t$ will
be mixtures
\begin{equation}
\begin{split}
\ket{{\Bz}_{\text{phys}}(t)} &= 
\bigl(g_+(t) + z\cdot g_-(t)\bigr)\,\ket{\Bz} -
\sqrt{1-z^2}\cdot \frac{q}{p}\,g_-(t)\,\ket{\Bzb} \\
\ket{{\Bzb}_{\text{phys}}(t)} &=
\bigl(g_+(t) - z\cdot g_-(t)\bigr)\,\ket{\Bzb} -
\sqrt{1-z^2}\cdot \frac{p}{q}\,g_-(t)\,\ket{\Bz}~,
\end{split}
\label{eq:general}
\end{equation}
where we have introduced
\begin{equation}
g_\pm(t)=\frac 12(e^{-i\omega_+t}\pm e^{-i\omega_-t})~.\label{eq:gdefn}
\end{equation}

At the $\Upsilon(4S)$ resonance, neutral \B\ mesons are produced in coherent pairs.  
If we subsequently observe a final state $f_1$ at time
$t_0=0$ and another state $f_2$ at some other time $ t$, either positive
or negative, we cannot in general know whether $f_1$ came from the
decay of a \Bz\ or a \Bzb\, and similarly for the state $f_2$.  If
$A_{1,2}$ and ${\overline A}_{1,2}$ are the amplitudes for the decay
of \Bz\ and \Bzb\ to the states $f_1$ and $f_2$, then the overall amplitude 
when $ t>0$ is given by 
\begin{eqnarray}
{\cal A}
&=&a_+g_+(t)+a_-g_-(t)~,
\end{eqnarray}
where
\begin{eqnarray}
a_+&=&-A_1{\overline A}_2+{\overline A}_1 A_2\nonumber \\
a_-&=&\sqrt{1-z^2}\left[{p\over q}A_1A_2-{q\over p}{\overline A}_1{\overline A}_2\right]
+z[A_1 {\overline A}_2+{\overline A}_1A_2]~.
\end{eqnarray}

Using the relations \begin{equation}
|g_\pm(t)|^2 = \frac{1}{2}\,e^{-t/\taub}\left[
\cosh(\Delta\Gamma t/2) \pm \cos(\Delta m t)\right]
\label{eq:gpmdefn}
\end{equation}
and
\begin{equation}
g_+^\ast(t)\,g_-(t) = -\frac{1}{2}\,e^{-t/\taub}\left[
\sinh(\Delta\Gamma t/2) + i\,\sin(\Delta m t)\right]~,
\label{eq:gpgmdefn}
\end{equation}
with $\taub=1/\Gamma$, we find the decay rate, which in fact is correct for $t$ positive or negative,
\begin{equation}
{{\rm d}N\over {\rm d}t}\propto e^{-\Gamma|t|}\Bigl\{{1\over 2} c_+\cosh(\Delta \Gamma t/2)
+{1\over 2}c_-\cos(\Delta m t) -\re \,s\, \sinh(\Delta \Gamma t/2)
+\im \,s\,\sin(\Delta m t) \Bigr\}\label{eq:pref}
\end{equation}
where
\begin{equation}
c_\pm=|a_+|^2\pm|a_-|^2~;\quad s=a_+^*a_-~.
\end{equation}

Now let us take $f_1 \equiv f_{tag}$ to be the state that is incompletely
reconstructed and which provides the tagging decay, and $f_2 \equiv f_{rec}$ the fully
reconstructed state (flavor or \CP\ eigenstate).  Because the
tagging algorithm is imperfect, we may incorrectly identify the flavor
of the decaying \B\ meson.  This can be accounted for by incoherently
combining correct and incorrect tags. A more subtle problem arises
because there may be a basic ambiguity:  the state $f_{tag}$ may
result from interference between decay from a \Bz\ and decay from a
\Bzb.  We consider first the simpler situation where there is no
underlying ambiguity.

If the tag is a \Bzb, we display this explicitly writing  ${\overline A}_{{\overline{tag}}} \neq 0, A_{{\overline{tag}}}=0$.  We define 
\begin{equation}
\lambda_{rec}=\frac qp \frac{\overline A_{rec}}{A_{rec}}\label{eq:lambda}~,
\end{equation}
which is independent of phase conventions for the \Bz\ and \Bzb\ states.
Dropping terms of order $z^2$, we find a decay rate

\begin{eqnarray}
{{\rm d}N\over {\rm d}t}{(tag=\Bzb)}&\!\propto\!&|{\overline A}_{\rm \overline{tag}}|^2|A_{rec}|^2 e^{-\Gamma|t|}\Biggl\{{1\over 2} \left[1+|\lambda_{rec} -z|^2\right]\cosh(\Delta \Gamma t/2)\nonumber\\
&&\qquad +{1\over 2} \left[1-|\lambda_{rec} -z|^2\right]\cos(\Delta m t)\nonumber\\
&&\qquad -\re (-\lambda_{rec} +z) \sinh(\Delta \Gamma t/2)
+\im (-\lambda_{rec}+z)\sin(\Delta m t)\Biggr\}~.\label{eq:bbartag_general}
\end{eqnarray}
Correspondingly, if $f_{tag}$ is tagged as a \Bz, ${\overline A}_{tag}=0,\ A_{tag}\neq 0$, and we have 

\begin{eqnarray}
{{\rm d}N\over {\rm d}t}{(tag=\Bz)}&\!\propto\!&|{A}_{\rm tag}|^2|A_{rec}|^2\left|\frac pq\right|^2 e^{-\Gamma|t|}\Biggl\{{1\over 2} \left[|\lambda_{rec}|^2+|1+z\lambda_{rec}|^2\right]\cosh(\Delta \Gamma t/2)\nonumber\\
&&\qquad +{1\over 2}  \left[|\lambda_{rec}|^2-|1+z\lambda_{rec}|^2\right]\cos(\Delta m t)\nonumber\\
&&\qquad +\re [\lambda_{rec}^*(1+\lambda_{rec} z)] \sinh(\Delta \Gamma t/2)
-\im [\lambda_{rec}^*(1+\lambda_{rec} z)]\sin(\Delta m t)\Biggr\}~.\nonumber\\
\label{eq:btag_general}
\end{eqnarray}
The normalizations are identical in Eqs. (\ref{eq:bbartag_general}) and 
(\ref{eq:btag_general}).

We consider several scenarios.
If $f_{rec}$ is a \CP\ eigenstate, then $|\lambcpbare|=\Rcp \absqop$,
with $\Rcp=|\overline A_{\CP}/A_{\CP}|$. If all the weak decay mechanisms 
have the same weak phase, $\Rcp=1$.
This is expected for final states like $\jpsi\KS$, where indeed measurements show
$|\lambcpbare|\approx 1$ \cite{ref:sin2b-babar}.  Dropping
quadratic terms in $z$ and $\dG$  we have

\begin{eqnarray}
{{\rm d}N\over {\rm d}t}{(tag=\Bzb; rec=\CP)}&\!\propto\!&|{\overline A}_{\rm \overline{tag}}|^2|A_{\CP}|^2 e^{-\Gamma|t|}\Biggl\{{1\over 2} \left[1+|\lambcpbare|^2 -2\re \lambcpbare \re z -2\im \lambcpbare \im z \right] \nonumber \\
&&\qquad+ {1\over 2} \left[1-|\lambcpbare|^2+2\re \lambcpbare \re z + 2\im \lambcpbare \im z \right]\cos(\Delta m t)\nonumber\\
&&\qquad+ \re \lambcpbare \sinh(\Delta \Gamma t/2)
-\im \left[\lambcpbare-z\right]\sin(\Delta m t)\Biggr\}~,\label{eq:bbartag_CP}
\end{eqnarray}

\begin{eqnarray}
{{\rm d}N\over {\rm d}t}{(tag=\Bz; rec={\CP})}&\!\propto\!&|{A}_{\rm tag}|^2|A_{\CP}|^2\left|\frac pq\right|^2 e^{-\Gamma|t|}\Biggl\{{1\over 2} \left[1+|\lambcpbare|^2+2\re \lambcpbare \re z - 2\im \lambcpbare \im z \right] \nonumber\\
&&\qquad +{1\over 2}  \left[|\lambcpbare|^2-1-2\re \lambcpbare \re z + 2\im \lambcpbare \im z \right]\cos(\Delta m t)\nonumber\\
&&\qquad +\re \lambcpbare \sinh(\Delta \Gamma t/2)
+\im \left[\lambcpbare-|\lambcpbare|^2z\right]\sin(\Delta m t)\Biggr\}~.\label{eq:btag_CPa}
\end{eqnarray}
Data from directly related final states like $\jpsi\KS$, with $\eta_{\CP}=-1$, and $\jpsi\KL$, with $\eta_{\CP}=+1$, 
where $\eta_{\CP}$ is the \CP\ eigenvalue of the final state,
can be combined by assuming that they are identical, except for an overall sign in \lambcpbare.

We assume that the decays of flavor eigenstates are dominated by a single weak mechanism, so that
$|A_{\rm flav}|=|{\overline A}_{{\overline {\rm flav}}}|$, 
$|{\overline{A}}_{\rm flav}|=|{A}_{{\overline {\rm flav}}}|$.  This will
enable us to relate the four possibilities that arise from the tag and reconstructed state being either \Bz\ or \Bzb.
When the fully reconstructed meson $f_{rec}$ is a flavor eigenstate,
$|\lambda_{rec}|$ is either very small or very large.  If it appears to come from
a \Bz, then $|\lambflavbare|\ll 1$ and we have 

\begin{eqnarray}
{{\rm d}N\over {\rm d}t}{(tag=\Bzb; rec=\Bz)}&\!\propto\!&|{A}_{\rm {tag}}|^2|A_{{\rm flav}}|^2 e^{-\Gamma|t|}
 \Biggl\{ {1\over 2}
+{1\over 2}\cos(\Delta m t)
+\im (-\lambflavbare+z)\sin(\Delta m t)\Biggr\}~,\nonumber\\ \label{eq:bbartag_breco}
\\
{{\rm d}N\over {\rm d}t}{(tag=\Bz; rec=\Bz)}&\!\propto\!&|{A}_{\rm tag}|^2|A_{\rm flav}|^2 \left|\frac pq\right|^2e^{-\Gamma|t|}\Biggl\{ {1\over 2}
-{1\over 2}\cos(\Delta m t)
+\im \lambflavbare\sin(\Delta m t)\Biggr\}~.\nonumber\\ \label{eq:btag_breco}
\end{eqnarray}
Conversely, if the fully reconstructed state is nominally a \Bzb,
$|\lambflavbarbare|\gg 1$ and

\begin{eqnarray}
{{\rm d}N\over {\rm d}t}{(tag=\Bzb; rec=\Bzb)}&\!\propto\!&|{A}_{\rm {tag}}|^2|{A}_{\rm \overline{flav}}|^2|\lambflavbarbare|^2 e^{-\Gamma|t|} \Biggl\{ {1\over 2}
-{1\over 2}\cos(\Delta m t)
+\im \lambbarflavbare\sin(\Delta m t)\Biggr\} \nonumber\\
&\!\propto\!&|{A}_{\rm{tag}}|^2|A_{\rm flav}|^2 \left|\frac qp\right|^2e^{-\Gamma|t|} \Biggl\{ {1\over 2}
-{1\over 2}\cos(\Delta m t)
+\im \lambbarflavbare\sin(\Delta m t) \Biggr\}~,\nonumber\\ 
\label{eq:bbartag_bbarreco}
\end{eqnarray}

\begin{eqnarray}
{{\rm d}N\over {\rm d}t}{(tag=\Bz; rec=\Bzb)}&\!\propto\!&|{A}_{\rm tag}|^2|{A}_{\rm \overline{flav}}|^2 \left|\frac pq\right|^2|\lambflavbarbare|^2e^{-\Gamma|t|} \Biggl\{ {1\over 2}
+{1\over 2}\cos(\Delta m t) \nonumber\\
& & \qquad -\im [z+\lambbarflavbare]\sin(\Delta m t) \Biggr\} \nonumber\\
&\!\propto\!&|{A}_{\rm tag}|^2|{A}_{\rm {flav}}|^2e^{-\Gamma|t|} \Biggl\{ {1\over 2}
+{1\over 2}\cos(\Delta m t)
-\im [z+\lambbarflavbare]\sin(\Delta m t)\Biggr\}~,\nonumber\\ 
\label{eq:btag_bbarreco}
\end{eqnarray}
where $\overline{\lambda}_{\rm flav} \equiv 1/\lambda_{\overline{\rm flav}}$.
  We see that the
overall normalization in the mixed final states has a factor of $|q/p|^2$ or
$|p/q|^2$ relative to the unmixed final states.

Doubly-\ckm-suppressed decays, such as $\Bz\to D^+\pi^-$, 
occur
at a rate roughly $|V_{ub}^*V_{cd}/V_{cb}^*V_{ud}|^2 \approx (0.02)^2$,
and can be ignored.  However, interference between favored and
suppressed decays are reduced by a factor of approximately $0.02$.  
For decays reconstructed in a final state as apparent $\Bz$ mesons, we anticipate
$|\lambflavbare |\approx 0.02 |q/p|$, while for reconstructed $\Bzb$ mesons,
$|\lambbarflavbare  |\approx 0.02 |p/q|$. 
In principle, every hadronic final
state has a different $\lambflavbare$ that can be written as
$\lambflavbare=|\lambflavbare |e^{-i\phi_s}e^{-i\phi_w}$, where
$\phi_s$ is the strong phase and $\phi_w$ is the weak phase.  Provided that
a single mechanism contributes to the allowed and suppressed decays,
$\lambflavbare$ and $\lambbarflavbare$ have the same
strong phase but opposite sign weak phase, and the magnitudes are the same
up to a relative $|p/q|^2$ factor.
For simplicity, in our 
analysis we take the \dckm\ effect to be equal for all reconstructed flavor states.

Eqs. (\ref{eq:bbartag_general})-(\ref{eq:btag_bbarreco}) show that
while \imlambcpbare, \imZ, \absqop\ and \Rcp\ are unambiguously 
determined, \reZ\ appears only in the product
\relambcpbare\reZ.  Similarly, \dG\ cannot be
determined separately from \relambcpbare\ since there is an ambiguity
in \relambcpbare: $\relambcpbare=\pm\sqrt{|\lambcpbare|^2-(\imlambcpbare)^2}$.  
As a result, the parameters which can actually be determined by
the analysis are \sgndGoverG, \absqop, \reZparflat, \imZ,
\imlambcpflat, \Rcp, \dM\ and $\taub=1/\Gamma$.
Both \CP\ eigenstates and flavor eigenstates are needed for the
analysis, as shown in Table~\ref{table:sensitivity}.
The sensitivity to \reZparflat\ and \imlambcpflat\ is
provided by the \cp-eigenstate events \bcp, for which the $t$ dependence is even
for the former and odd for the latter. The \bflav\ sample contributes
marginally because it lacks explicit dependence on \imlambcpflat\ and the
dependence with \reZ\ is scaled by the $\sinh\left( \dG t / 2 \right)$ term, 
which is small for small \dG. In contrast, \absqop\ and \imZ\ (and
\dM) are completely dominated by the large statistics \bflav\ sample,
for which the $t$ dependence is even for the former and odd for the
latter. For small values of \dGoverG, the determination of \dGoverG\
is dominated by the \bcp\ sample, in spite of the relatively small
statistics compared to the \bflav\ sample. This is due to the even
$\cosh(\dg t /2)$ dependence ($\dg^2$ to first order) of the flavor
sample, while the \CP\ sample has a non-vanishing odd $\sinh(\dg t/2)$
($\dg$ to first order) dependence.  The contribution of  $\sinh(\dg t/2)$
is the same for both \Bz\ and \Bzb\ tags, so untagged events may be included
as well. The \bcp\ sample is also sensitive to the sign of \dGoverG\ (up to the
sign ambiguity from \relambcpbare). Overall, the combined use of the
\bflav\ and \bcp\ samples provides maximal sensitivity to the physical
parameters, since they are determined either from different samples,
or from different $t$ dependencies. Small correlations are induced by
the detector resolution.

\begin{table}\begin{center}
\caption[a]{Dominant sensitivity of physical parameters to the distributions
measured with the fully reconstructed flavor and \CP\ states.  The flavor sample
is much larger than the \CP\ sample.}\label{table:sensitivity}
\vspace{0.1in}
\begin{tabular}{lcccc}\\
&\multicolumn{2}{c}{\bflav}&\multicolumn{2}{c}{\bcp}\\ 
Variable&$t$-even&$t$-odd&$t$-even&$t$-odd\\ \hline\hline
$\sgndGoverG$&&&&$\times$\\ 
$\absqop$&$\times$&$\times$&&\\ 
$\reZparflat$&&&$\times$& \\ 
\imZ&&$\times$&&\\ 
$\imlambcpflat$&&&&$\times$\\ 
$\Rcp$&&&$\times$&\\ 
\dM & $\times$ &&&\\ \hline
\end{tabular}\end{center}
\end{table}



Doubly-\ckm-suppressed decays occur on the tagging side as well, 
if the decay is non-leptonic.  Consider, for example, the contribution
of doubly-\ckm-suppressed decays when the tagging decay is ostensibly
a \Bz.  To first order in $\lambda_{\rm tag}=(q/p){\overline A}_{\rm tag}/A_{\rm tag}$, $z$ and \dGoverG,
the new contribution to the decay rate is 

\begin{eqnarray}\lefteqn{
{{\rm d}N\over {\rm d}t}{(tag=\Bz;\ \dckm\ {\rm contribution})}\propto  |{A}_{\rm tag}|^2|A_{rec}|^2\left|\frac pq\right|^2 e^{-\Gamma|t|}} \\
&&\times\Biggl\{{1\over 2} \left[-4\re  \lambda_{\rm tag}\re \lambda_{rec}\right]
+{1\over 2}  \left[-4\im  \lambda_{\rm tag}\im \lambda_{rec}\right]\cos(\Delta m t)
+\im \lambda_{\rm tag}(|\lambda_{rec}|^2-1)\sin(\Delta m t)\Biggr\}~. \nn \label{eq:btag_CPb}
\end{eqnarray}
This shows that if the reconstructed state is a flavor eigenstate, the
\dckm\ effect in tagging is negligible except in the $\sin(\Delta m t)$
term. Conversely, if the reconstructed state is a \CP\ eigenstate with
$|\lambda_{rec}| \approx 1$, the \dckm\ effect is confined to the terms even in
$t$. Because terms quadratic in $\lambda_{\rm tag}$ can be ignored, the
combined effect of \dckm\ on tagging can be incorporated in one value of
$\lambda_{\rm tag}$ and one of $\overline{\lambda}_{\rm tag} \equiv 1/ \lambda_{\rm \overline{tag}}$.  


\section{\boldmath The \babar\ detector}
\label{sec:detector}
The data used in this analysis were collected with the \babar\
detector at the \pep2\ storage ring.  The \babar\ detector is
described in detail elsewhere~\cite{ref:babar-nim}, so here we provide
only a brief description of the apparatus.  

Surrounding the beam-pipe is a five-layer silicon vertex tracker (SVT),
which provides precise measurements of points along the trajectories of charged
particles as they leave the interaction region. This allows track
reconstruction, even for some particles with momentum less than
$120\mevc$.  Outside of the SVT is a 40-layer drift chamber (DCH)
filled with an 80:20 helium-isobutane gas mixture, chosen to minimize
multiple scattering.  The DCH measurements provide charged-particle tracking
and determination of momenta through track curvature in 
the 1.5-T magnetic field generated by the superconducting coil.  
The DCH also provides $dE/dx$ energy-loss
measurements, which contribute to charged-particle identification.
Surrounding the drift chamber is a novel detector of internally
reflected Cerenkov radiation (DIRC), which provides charged-particle
identification in the barrel region. Outside of the DIRC is a CsI(Tl)
highly segmented electromagnetic calorimeter (EMC), which is used to
measure the energy of photons, to provide electron identification, and
to detect neutral hadrons through shower shapes.
Finally, the flux return of the superconducting coil surrounding the
EMC is instrumented with resistive plate chambers interspersed with
iron (IFR) for the identification of muons and neutral hadrons.

A detailed Monte Carlo program based on the GEANT4~\cite{ref:geant4}
software package is used to simulate the \babar\ detector response and
performance. The agreement between data and simulation is very good
\cite{ref:babar-nim}.



\section{\boldmath Data samples and \B\ meson reconstruction}
\label{sec:sample}

We have selected
events where one of the \B\ mesons is completely reconstructed in either a flavor (\bflav) 
or \CP\ (\bcp) eigenstate, using the same criteria used for the 
\babar\ hadronic \dM\ \cite{ref:dM-babar-had,ref:babar-stwob-prd} and 
$\sintwob$ measurements~\cite{ref:sin2b-babar}.
The decay modes used for the flavor sample, the \CP\ sample, and a control
sample are displayed in Table~\ref{table:modes}.
\begin{table}[h!]
\begin{center}
\caption{The flavor, \CP, and control samples used in this analysis. Charged and 
neutral flavor eigenstate decay modes imply also their charge conjugate.}\label{table:modes}  
\vspace{0.1in}
\begin{tabular}{lll}\hline 
Samples & Decay modes & \\
\hline \hline
$\bflav$&&\\
&$\Bz \to D^{*-}\pi^+(\rho^+,a_1^+)$&\\
&&$D^{*-}\to {\overline D}^0\pim$\\
&&$ {\overline D}^0\to \Kp\pim, \Kp\pim\piz, K^+\pim\pip\pim, \KS\pip\pim$\\
&&$\rho^+\to \pip\piz$\\
&&$a_1^+\to \pip\pip\pim$\\ 
&$\Bz \to D^{-}\pi^+(\rho^+,a_1^+)$&\\
&&$D^{-}\to \Kp\pim\pim, \KS\pim$\\
&$\Bz \to \jpsi \Kstarz$&\\
&&$\Kstarz \to \Kp \pim$\\ \hline
$\bcp$&&\\
&$\Bz\to \jpsi\KS$&\\
&&$\jpsi \to e^+e^-,\mu^+\mu^-$\\
&&$\KS\to \pip\pim, \piz\piz$\\
&$\Bz\to \psitwos\KS$&\\
&&$\psitwos\to e^+e^-,\mu^+\mu^-,\jpsi\, \pip\pim$\\
&$\Bz\to \chic1\KS$&\\
&&$\KS\to \pip\pim$\\
&&$\chic1\to \jpsi\,\gamma$\\
&&$\KS\to \pip\pim$\\
&$\Bz\to \psi\KL$&\\ \hline
Control&&\\
&$\Bu \to \overline{D}^{(*)0}\pip$&\\
& & $\overline{D}^{*0} \to \overline{D}^{0} \piz$ \\
&$\Bu \to \jpsi \Kp$&\\
&$\Bu \to \psitwos \Kp$&\\
&$\Bu \to \chic1 \Kp$&\\
&$\Bu \to \jpsi K^{*+}$&\\
&&$ K^{*+}\to \KS\pip$\\ \hline
\end{tabular}
\end{center}
\end{table}

We select \bflav\ and \bcp\ candidates by requiring that the
difference $\Delta E$ between their energy and the beam energy in the
center-of-mass frame be less than three standard deviations from zero.
The \de\ resolution ranges between 10 and 50 \mev\ depending on the decay mode.
For \bflav\ and \bcp\ modes involving \KS\ (\bcpks), the beam-energy
substituted mass must be greater than $5.2\gevcc$.  The beam-energy substituted mass is given by
$\mes = \sqrt{(\frac 12 s + \pvec_i\cdot \pvec_{\B})^2/E_i^2 - p_{\B}^2}$, where $s$
is the square of the center-of-mass energy, $E_i$ and $\pvec_i$ are the
total energy and the three-momentum of the initial state in the laboratory frame, 
and $\pvec_{\B}$ is the three-momentum of the \B\ candidate in the same frame. 
In the case of decays to $\jpsi\KL$, the \KL\ direction is measured 
 but its momentum is only inferred by constraining the mass of the 
 $\jpsi\KL$ candidate to the known \Bz\ mass. 
As a consequence there is only one
parameter left to define the signal region, which is taken to be $|\Delta E| < 10\mev$. 

\begin{figure}[h!]
\begin{center}   
\epsfig{file=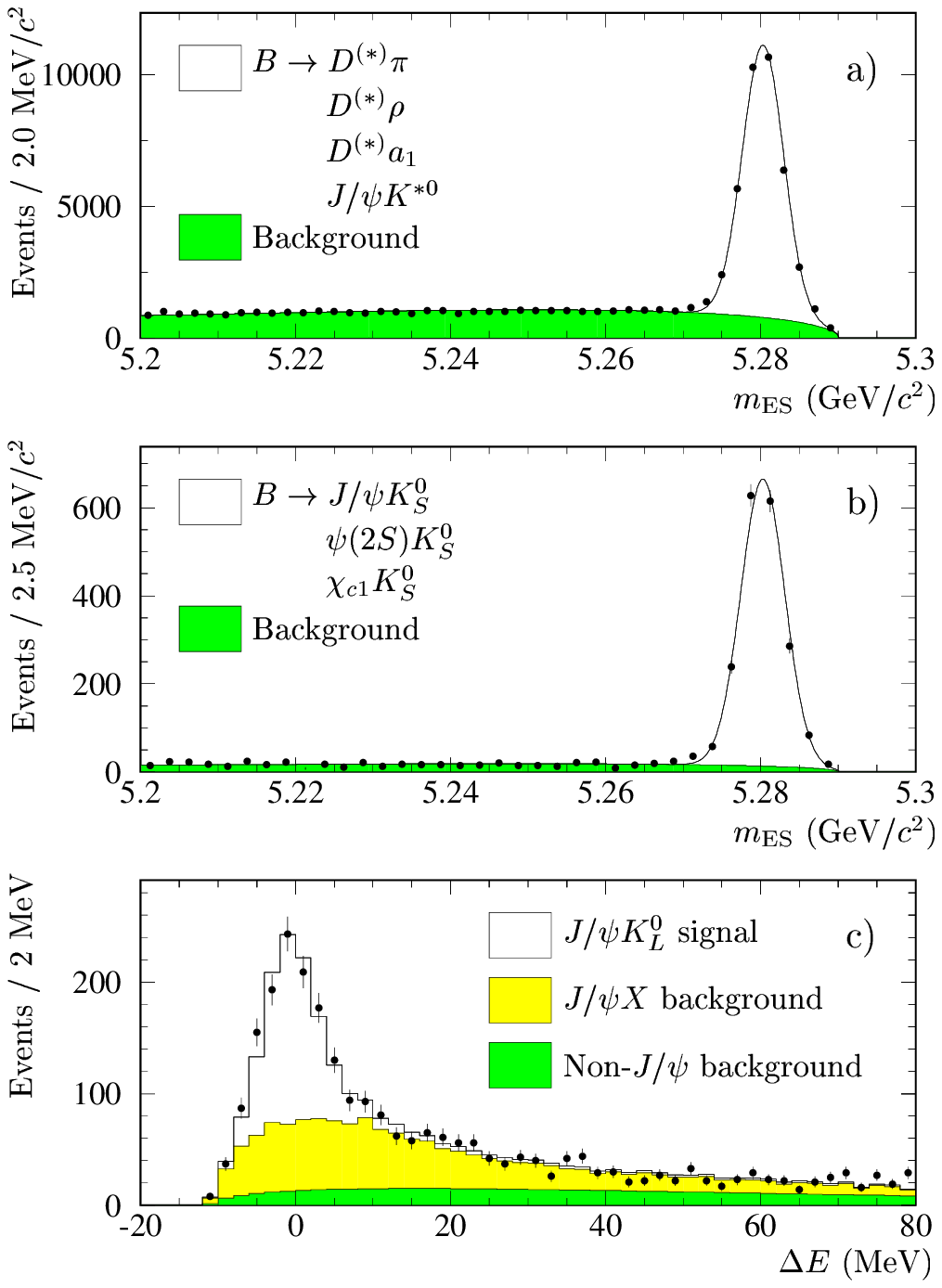,width=0.7\linewidth} \\
\end {center} 
   \caption{Distributions for \bflav\ and \bcp\ candidates before vertexing requirements: a) \mes\ for
\bflav\ states; b) \mes\ for $\Bz \to \jpsi \KS, \psitwos \KS, \chic1 \KS$ final states; and c)
$\Delta E$ for the final state $\Bz \to \jpsi \KL$.\label{fig:sample}}
\end{figure}

The
purities are determined from fitting the data to the \mes\ (\bflav\
and \bcpks\ modes) or $\Delta E$ (\bcpkl\ mode) distributions
\cite{ref:babar-stwob-prd}.  Figure \ref{fig:sample} shows the \mes\
distribution for the \bflav\ and \bcpks\ samples and the $\Delta E$
distribution for the \bcpkl\ candidates, before the vertexing
requirements (see Sec.~\ref{sec:decaytime}).
The combinatorial background in the \mes\ distributions 
 is described by an empirical phase-space model \cite{ref:babar-stwob-prd} and the signal 
 with a Gaussian distribution. The combinatorial background consists of 
 continuum and \BB\ sources, and has a time structure with both prompt 
 and non-prompt components. A small correlated background due to other 
 \B\ decays (not shown) also peaks at the \B\ mass. The background in 
 the $\jpsi\KL$ channel receives contributions from other \B\ decays 
 with real \jpsi\ mesons in the final state, and combinatorial sources.

After completely reconstructing one \B\ meson, the rest of the event
is analyzed to identify the flavor of the opposite \B\ and to
reconstruct its decay point, as described in Secs. \ref{sec:tagging}
and \ref{sec:decaytime}.

Using exactly the same requirements, we analyze GEANT4-simulated 
 samples of generic \BB\ and signal events to check for any biases 
 in the procedure or extracted parameters.
The Monte Carlo samples are also used to
assess detector response and to estimate some
background sources. The values of the physics parameters assumed in
the simulations are similar to those measured in the data. We used additional
samples with significantly different values to
check the reliability of the analysis in other regions of the
parameter space.

\section{\boldmath Flavor tagging}
\label{sec:tagging}

The tracks that are not part of the fully reconstructed \B\ meson are
used to determine whether the \btag\ was a \Bz\ or \Bzb\ when it
decayed.  This determination cannot be done perfectly.  If the probability of
an incorrect assignment is $w$,  an asymmetry that depends on the 
difference between \Bz\ and \Bzb\ tags will be reduced by a factor $D=1-2w$,
frequently called the dilution.  
A neural network combining the outputs of physics-based
algorithms is used to take into account the correlations
between the different sources of flavor information and to assign
the event to one of five mutually exclusive tagging categories.  The dilution
for each category is determined from data.
Grouping tags into several categories, each with a relatively narrow
range in mistag probability, increases the overall power of the tagging.

We group together events of similar character to make it possible to
study systematic effects. 
Events with an identified primary electron or muon and 
a supporting kaon, if present, are assigned to the {\tt Lepton} category.
The {\tt KaonI} category contains events with an identified kaon and a soft-pion candidate
with opposite charges and similar flight direction.
Soft pions from \Dstarp\ decays are selected on the basis of
their momentum and direction with respect to the thrust axis of
\btag. Events with only an identified kaon are assigned to the {\tt KaonI} or 
{\tt KaonII} category depending on the estimated mistag probability.
Events with only a soft-pion candidate are assigned to the {\tt KaonII} 
category as well. The remaining events are assigned to the {\tt
Inclusive} or {\tt UnTagged} category based on the estimated mistag
probability. The {\tt UnTagged} tagging category has a mistag rate near
50\%, and therefore does not provide tagging information but it increases
the sensitivity to the lifetime difference and allows the determination
from the data of the detector charge asymmetries, as described in
Sec.~\ref{sec:method}.  The tagging efficiencies $t^{\alpha}$, defined
after vertexing cuts, for the five categories are measured from the data and summarized in Table
\ref{tab:mistag} of Sec.~\ref{sec:results}. This tagging algorithm is identical 
to that used in Ref.~\cite{ref:sin2b-babar}.

The mistag probabilities appear separately  for \Bz\ and \Bzb\ tags in each tagging category. 
If we define $\wa = (\wa_{\Bz} + \wa_{\Bzb})/2$ and $\dwa = \wa_{\Bz} - \wa_{\Bzb}$, we have that
\begin{equation}
  \wa_{\BzBzb} = \wa \pm \dwa/2~.
\label{eq:wb0b0b}
\end{equation}
A correlation between the mistag rate and the \dt\
uncertainty estimated event-by-event (discussed in
Sec.~\ref{sec:decaytime}) is observed in the Monte Carlo simulation
for kaon based tags \cite{ref:dM-babar-dstlnu,ref:babar-stwob-prd}. 
For a \deltat\ uncertainty less than \mbox{1.4~\ps,} this correlation is found to be approximately
linear:

\begin{equation}
  \wa = \wa_0 + \wa_{\rm slope} \sdt~.
\label{eq:tagvtxcorr}
\end{equation}
All signal mistag parameters, $\wa_0$, $\wa_{\rm slope}$ and \dwa, are floated in the global fit 
(11 in total since $w^{\tt Lepton}_{\rm slope}$ is assumed to be zero), and their results can 
be found in Table \ref{tab:mistag} in Sec.~\ref{sec:results}.

\section{Decay time measurement and \dt\ resolution function}
\label{sec:decaytime}

The time interval 
$\dt=t_{\rm rec} - t_{\rm tag}$ 
between the two \B\ decays is calculated from
the measured separation \deltaz\ between the decay vertex of the
reconstructed \brec\ meson  and the vertex of the \btag\ meson along the 
$z$-axis, using the known boost of the \FourS\ resonance in the laboratory, $\beta\gamma=0.56$. The
method is the same as described in Sec.~V in Ref.~\cite{ref:babar-stwob-prd}.



An estimated error $\sigma_{\dt}$  on \dt\  is
calculated for each event. This error accounts for uncertainties in the track
parameters from the SVT and DCH hit resolution and multiple
scattering, our knowledge of the beam spot size and most of the
effects from the average \B\ flight length in the vertical
direction as well as the \deltaz\ to \dt\ conversion.
However, it does not account for errors due to mistakes of the
pattern recognition system, wrong associations of tracks to vertices,
misalignment within and between the tracking devices, inaccuracies in
the modeling of the amount of material in the tracking detectors and
in our knowledge of the beam spot position and size, and the absolute
$z$ scale. We use parameters in the
\dt\ resolution function, extracted from the data in the global 
fit (Table \ref{tab:resolution} in Sec.~\ref{sec:results}), 
to absorb most of these effects. Remaining systematic 
uncertainties are discussed in detail in Sec.~\ref{sec:systematics}.


We use only those events in which the vertices of the \brec\ and
\btag\ are successfully reconstructed and for which $| \dt | <
20$ ps and the \dt\ uncertainty ($\sdt$) is less than \mbox{1.4~ps,} as used in Ref.~\cite{ref:dM-babar-had}.  
The fraction of events in data satisfying these requirements is about
85\%.  From Monte Carlo simulation we find that the reconstruction
efficiency does not depend on the true value of \dt.  Excluding 0.3\%
of the events that are poorly measured, the rms
vertex resolution is about 160\mum\ (1.0 ps).
Because this resolution is poorer than that for the completely reconstructed
vertex,  the overall \deltaz\ resolution is dominated by the
resolution of the \btag\ vertex.  

To model the \dt\ resolution we use the sum of three Gaussian
distributions (called {\em core}, {\em tail} and {\em outlier}
components) with different means and widths:

\bea
{\cal R}(\delta t,\sdt;\vec{q}) & = &
(1-f_{\rm tail}-f_{\rm out})
               h_{\rm G}(\delta t;\delta_{\rm core}\sdt,S_{\rm core}\sdt) +  \nn \\
  & & f_{\rm tail} h_{\rm G}(\delta t;\delta_{\rm tail}\sdt,S_{\rm tail}\sdt) + \nn \\
  & & f_{\rm out} h_{\rm G}(\delta t;\delta_{\rm out},\sigma_{\rm out}) 
\label{eq:resol}
\eea
with
\bea
  h_{\rm G}(\delta t;\delta,\sigma) & = & 
       \frac{1}{\sqrt{2\pi}\sigma} \exp( -(\delta t-\delta)^2/(2\sigma^2)  )~,
\label{eq:resolGauss}
\eea
where $\vec{q}$ represents the collection of parameters needed to describe
the Gaussians and \mbox{$\delta t = \dt-\dt_{\rm true}$} represents the reconstruction
error.  The vertex reconstruction provides an event-by-event estimate for
$|\delta t|$, namely
$\sdt$. We incorporate the last Gaussian in Eq. (\ref{eq:resol}) without
reference to 
$\sdt$ since the outlier component is not expected to be well
described by the estimated error.  The first two Gaussian components
allow two independent scale factors, $S_{\rm core}$ and $S_{\rm tail}$, to
accommodate an overall underestimate ($>1$) or overestimate ($<1$) of
the estimated errors. The core and tail Gaussian distributions are
allowed to have non-zero means ($\delta_{\rm core}$ and $\delta_{\rm tail}$, respectively) 
to account for residual biases due to
daughters of long-lived charm particles included in the $B_{\rm tag}$
vertex.  Separate means are used for the core distribution of each tagging category.  These means are scaled by $\sdt$ to account for a correlation
observed in Monte Carlo simulation between the mean of the $\delta t$ distribution 
and $\sdt$ \cite{ref:dM-babar-dstlnu,ref:babar-stwob-prd}.
This correlation is found to be approximately linear for a $\sigma_{\dt}$ less than \mbox{1.4~ps.} 
The non-zero means of the resolution functions introduce an asymmetry
into the otherwise symmetric \dt\ distributions.  The outlier Gaussian
has a global width ($\sigma_{\rm out}$) and offset ($\delta_{\rm
out}$), and it accounts for less than 0.3\% of the reconstructed
vertices. The parameters $f_{\rm tail}$ and $f_{\rm out}$ in
Eq. (\ref{eq:resol}) represent, respectively, the fractions of the
tail and the outlier Gaussian components.  In simulated events, we
find no significant differences between the
\dt\ resolution function of the \bflav, \bcpks\ and \bcpkl\ samples.
This is expected, since the $B_{\rm tag}$ vertex
precision dominates the \deltat\ resolution. Hence, the same resolution
function is used for all modes. Residual differences are taken into account
in the evaluation of systematic errors, as described in Sec.~\ref{sec:systematics}.

We find that the three parameters
describing the outlier Gaussian component are largely correlated among
themselves and with other resolution function parameters. Therefore, we fix
the outlier bias $\delta_{\rm out}$ and width $\sigma_{\rm out}$ to 0 and 8
ps, respectively, and vary them in a wide range to evaluate systematic
uncertainties. The resulting signal resolution function is
described by a total of 12 parameters, $$\vec{q}=\left\{ S_{\rm core},
\delta_{\rm core}^{\tt Lepton}, \delta_{\rm core}^{\tt Kaon I},
\delta_{\rm core}^{\tt Kaon II}, \delta_{\rm core}^{\tt Inclusive},
\delta_{\rm core}^{\tt UnTagged}, f_{\rm tail}, \delta_{\rm tail},
S_{\rm tail}, f_{\rm out}, \delta_{\rm out}, \sigma_{\rm out}
\right\},$$ 10 of which are floated in the final fit.  Their final values
can be found in Table \ref{tab:resolution} in Sec.~\ref{sec:results}.

As a cross-check, we use an alternative resolution function that is
the sum of a single Gaussian distribution (centered at zero), the
same Gaussian convolved with a one-sided exponential to describe the
core and tail parts of the resolution function, and a single Gaussian
distribution to describe the outlier component
\cite{ref:dM-babar-dstlnu}. The exponential
component is used to accommodate the bias due to tracks from charm
decays on the \btag\ side.  The exponential constant is scaled by
$\sdt$ to account for the previously described correlation between the
mean of the $\delta t$ distribution and $\sdt$. In this case, each
tagging category has a different core component fraction and
exponential constant.

\section{\boldmath Likelihood fit method}
\label{sec:method}

We perform a single unbinned maximum-likelihood fit to all \bflav,
\bcpks\ and \bcpkl\ events.  Each event is characterized by its
assigned tag category, $\alpha\in \{\tt Lepton,~KaonI,~KaonII,~Inclusive,$ ${\tt UnTagged \} }$; its tag-flavor type, 
$tag=\Bz, \Bzb$
(unless it is untagged); its reconstructed event type, 
$rec=\Bz, \Bzb, \cpks, \cpkl$; 
the values of $\dt$ and $\sdt$; and a
variable $\zeta$, either \mes\ or \de, used to
assign the probabilities that the event is signal or background.
The underlying distributions depend on whether the event is signal
or any of a variety of backgrounds
(together specified by $j$),
on the tag category ($\alpha$), on the tag flavor
($tag$), and on the reconstructed final state type ($rec$). The contribution
of a single event to the log-likelihood is

\bea
\ln \left[ \sum_j {\cal F}^{\alpha,j}_{rec}(\zeta) {\cal H}_{tag,rec}^{\alpha,j}(\dt,\sdt) \right]~.
\label{eq:lik}
\eea
For a given reconstructed event type $rec$ and tagging category
$\alpha$, $ {\cal F}^{\alpha,j}_{rec}(\zeta)$ gives the probabilities
that the event belongs to the signal or any of the various backgrounds
indicated by $j$. Each such combination has its own PDF $ {\cal
H}_{tag,rec}^{\alpha,j}(\dt,\sdt)$, which depends as well on the
particular tag flavor, $tag$.  This distribution is a convolution of a
tagging-category-dependent time distribution, 
${ H}^{\alpha,j}_{tag,rec}(\dt_{\rm true})$, with a \dt\ resolution function

\bea
  {\cal H}^{\alpha,j}_{tag,rec}(\dt,\sdt) & = & \int_{-\infty}^{+\infty} {\rm d}(\dt_{\rm true}) {\cal
  R}(\dt-\dt_{\rm true},\sigma_{\dt};\vec{q}^{~\alpha,j}) { H}^{\alpha,j}_{tag,rec}(\dt_{\rm true}) 
\label{eq:intensitiesAnaResol}
\eea
where
\begin{eqnarray}
  { H}^{\alpha,j}_{tag,rec}(\dt_{\rm true}) & = & r_{rec}^j \left\{ 
       t^{\alpha,j}_{tag} (1 - \waj_{tag}) { h}^{ j}_{tag,rec}(\dt_{\rm true})\right. \nonumber\\
&&\qquad +\left.  
       t^{\alpha,j}_{{\overline{tag}}} \waj_{{\overline{tag}}} { h}^{ j}_{{\overline{tag}},rec}(\dt_{\rm true})
        \right\}~.
\label{eq:intenmistag}
\end{eqnarray}
Here ${ h}^j_{tag,rec}(\dt_{\rm true})$ represents the time
dependence ${{\rm d}N / {\rm d} \dt_{\rm true}}$ described in Sec.~\ref{sec:decayrates}, with 
\mbox{$\dt_{\rm true} \equiv t$}, 
using full expressions rather than expansions in \dG\ and $z$. We indicate by
$\waj_{tag/{\overline{tag}}}$ the mistag fractions for category
$\alpha$ and component $j$. The index ${\overline{tag}}$ denotes the
opposite flavor to that given by $tag$. For events falling into
tagging category {\tt UnTagged},
$\waj_{tag/{\overline{tag}}}=1/2$. The efficiency $t^{\alpha,j}_{tag}$ is the probability that an event whose signal/background
nature is $j$ and whose true tag flavor is $tag$ will be assigned to
category $\alpha$, regardless of whether the flavor assigned is
correct or not.  The efficiency $r^j_{rec}$ is the probability
that an event whose signal/background nature is indicated by $j$ and
whose true character is $rec$ will, in fact, be reconstructed.

\subsection{PDF normalization}

Every reconstructed event, whether signal or background occurs at some
time $\Delta t_{\rm true}$, so

\bea
\int_{-\infty}^{+\infty}\, {\rm d}(\Delta t_{\rm true})
    {h}^{j}_{tag,rec}(\dt_{\rm true}) & = & 1 ~,~~~
{\rm for~all}~rec,tag~{\rm and}~j~.
\label{eq:norm}
\eea
Moreover, every event is assigned to some tag category (possibly {\tt UnTagged}), thus
\begin{equation}
\sum_\alpha t^{\alpha,j}_{tag}=1~.
\end{equation}
It follows then that the normalization of  $ {H}^{\alpha,j}_{tag,rec}(\dt_{\rm true})$ is
\begin{equation}
\sum_\alpha\sum_{tag} \int_{-\infty}^{+\infty}\, {\rm d}(\Delta t_{\rm true})
    {H}^{\alpha,j}_{tag,rec}(\dt_{\rm true})=r^j_{rec}~.
\label{eq:norm2}
\end{equation}
In this analysis the nominal normalization of ${\cal H}^{\alpha,j}_{tag,rec}(\dt,\sdt)$ is the same
as $ {H}^{\alpha,j}_{tag,rec}(\dt_{\rm true})$ (asymptotic normalization), but
fits with normalization in the interval $[-20,20]$ ps were also performed as a cross-check to evaluate 
possible systematic effects.

\subsection{Efficiency asymmetries}

For each signal or background, $j$, there is an average reconstruction efficiency,
$r^j = (r^j_{\Bz}+r^j_{\Bzb})/2$.  These average efficiencies are ultimately
absorbed when we define fractions of reconstructed events falling into the different
signal and background classes.  In contrast, because all events fall into some tagging
category (including {\tt UnTagged}), the averages $t^{\alpha,j} = (t^{\alpha,j}_{\Bz}+
t^{\alpha,j}_{\Bzb})/2$ are meaningful, 
and for $j$=signal, and $\alpha={\tt UnTagged}$ the result plays an important role.
The asymmetries in the efficiencies, 
\begin{eqnarray}
\nu^j&=&\frac{r^j_{\Bz}-r^j_{\Bzb}}{r^j_{\Bz}+r^j_{\Bzb}}~,\nonumber\\
\mu^{\alpha,j}&=&\frac{t^{\alpha,j}_{\Bz}-t^{\alpha,j}_{\Bzb}}{t^{\alpha,j}_{\Bz}+
    t^{\alpha,j}_{\Bzb}}~,
\label{eq:numudef}
\end{eqnarray}
need to be determined because they might otherwise mimic fundamental
asymmetries we seek to measure. In Appendix \ref{appendix:asymmetries}
we illustrate how the use of the untagged sample makes it possible to determine
the asymmetries in the efficiencies. Note that asymmetries due to differences
in the magnitudes of the decay amplitudes, $|A_{\flavz}| \ne |\overline{A}_{\flavzb}|$ 
and $|A_{\tagz}| \ne |\overline{A}_{\tagzb}|$, 
cannot be distinguished from the asymmetries in the efficiencies, 
thus are absorbed in the $\nu$, $\mu^{\alpha}$ parameters.

We fix the average tagging efficiencies $t^{\alpha,j}$ to estimates
determined by counting the number of events falling into the different
tagging categories, for each decay channel separately and in all the
$\zeta$ range, after vertexing cuts.  The parameters $\nu^{\rm sig}$
and $\mu^{\alpha,\rm sig}$ (signal events) are included as free parameters in
the global fit, and are assumed to be the same for all \Bz\ peaking
background sources.
For \Bu\ peaking
background components, the $\nu$s and $\mu$s are fixed to the values extracted from a
previous unbinned maximum-likelihood fit to the tagged and untagged \dt\
distributions of \Bu\ data used as control samples, described in
Sec.~\ref{sec:sample}. We neglected $\nu^{j}$ and $\mu^{\alpha,j}$ for
combinatorial background sources.

\subsection{Signal and background characterization }

The function ${\cal F}^{\alpha,j}_{rec}(\zeta)$ in Eq. (\ref{eq:lik})
describes the signal or background probability of observing a
particular value of $\zeta$. It satisfies
\begin{equation}
\int_{\zeta_{\rm min}}^{\zeta_{\rm max}} {\rm d}\zeta \sum_j {\cal F}^{\alpha,j}_{rec} (\zeta)  =  1~,
\end{equation}
where $[\zeta_{\rm min},\zeta_{\rm max}]$ is the range of \mes/\de\ values
used for analysis.  

For \bflav\ and \bcpks\ events, the \mes\ shape is described with a single Gaussian distribution for
the signal and an ARGUS parameterization for the background
\cite{ref:babar-stwob-prd}. Based on these fits, an event-by-event
signal probability $p^{\alpha,\rm sig}_{rec}(\mes)$ can be calculated for
each tagging category $\alpha$ and sample $rec$. As we do not expect
signal probability differences between
\Bz\ and \Bzb, the \mes\ fits were performed to \bflavz\ and \bflavzb\ events together. Due to the lack of
statistics and the high purity of the sample, the \mes\ fits to the
$\Bz \to \psitwos\KS$ and $\Bz \to \chicone\KS$ samples were performed
without splitting by tagging category. The component fractions ${\cal
F}^{\alpha,j}_{rec}(\mes)$ are then given by

\bea
{\cal F}^{\alpha,\rm sig}_{rec}(\mes) & = &
\left[ 1-f^{\alpha,\rm peak}_{rec} \right] p^{\alpha,\rm sig}_{rec}(\mes) \nn \\
{\cal F}^{\alpha,\rm peak}_{rec}(\mes) & = &
   f^{\alpha,\rm peak}_{rec} p^{\alpha,\rm sig}_{rec}(\mes) \nn \\
{\cal F}^{\alpha,\beta}_{rec}(\mes) & = &
\left[ 1 - p^{\alpha,\rm sig}_{rec}(\mes) \right] f^{\alpha,\beta}_{rec}
\eea
where $\beta$ indexes the combinatorial, prompt and non-prompt, background components, 
\begin{equation}
\sum_\beta  f^{\alpha,\beta}_{rec} =1~.
\end{equation}
The fraction $f^{\alpha,\rm peak}_{rec}$ of the signal Gaussian
distribution is due to backgrounds that peak in the same region as the signal, determined from Monte
Carlo simulation \cite{ref:babar-stwob-prd}. The estimated
contributions are $(1.5\pm0.6)\%$, $(0.28\pm0.11)\%$, $(1.8\pm0.6)\%$,
$(1\pm3)\%$ and $(3.5\pm1.4)\%$ for the \bflav, $\jpsi\KS\ (\KS \to
\pipi)$, $\jpsi\KS\ (\KS \to \ppz)$, $\psitwos\KS$ and $\chicone\KS$
channels, respectively. A common peaking background fraction is
assumed for all tagging categories within each decay mode.  
We take a common prompt fraction for all tagging categories for each \bcpks\
decay channel independently.  Due to the higher statistics of the
\bflav\ sample and the significant differences in the background
levels for each tagging category, $f^{\alpha,\rm prompt}_{\Bz}=f^{\alpha,\rm prompt}_{\Bzb}$ is
allowed to vary with the tagging category.  Note that the parameters
of the ${\cal F}^{\alpha,\rm sig}_{rec}(\mes)$ functions, determined from
the set of separate unbinned maximum-likelihood fits to the \mes\
distributions, are kept fixed in the global fit.

For \bcpkl\ events the background level is much
higher, with significant non-combinatorial components, therefore
requiring special treatment \cite{ref:babar-stwob-prd}.  A binned
likelihood fit to the \de\ spectrum in the data is used to determine
the relative amounts of signal and background from $\B \to \jpsi X$
events and events from a misreconstructed $\jpsi \to \ell^+ \ell^-$
candidate (non-\jpsi\ background). In these fits, the signal and $\B
\to \jpsi X$ background distributions are obtained from inclusive
\jpsi\ Monte Carlo, while the non-\jpsi\ distribution is obtained from
the \jpsi\ dilepton mass sideband.  The Monte Carlo simulation is also
used to evaluate the channels that contribute to the $\B \to \jpsi X$
background. Due to differences in purity and background composition,
the fit is performed separately for each \KL\ reconstruction type (EMC
and IFR) and lepton type (\jpsi \to \epem\ and \jpsi \to \mumu).
The different inclusive \jpsi\ backgrounds from Monte Carlo are then renormalized to the \jpsi\ background fraction 
extracted from the data. The fractions are adjusted for lepton-tagged and non-lepton-tagged events in order to
compensate for the observed differences in flavor tagging efficiencies in the \jpsi\ sideband events relative to 
the \bflav\ and inclusive \jpsi\ Monte Carlo. 
In addition, some of the decay modes in the inclusive \jpsi\ background have
particular \CP\ content. The PDF can then be formulated as

\bea
  \sum_j {\cal F}_{\cpkl}^{\alpha,j}(\de) {\cal H}^{\alpha,j}_{tag,\cpkl}(\dt,\sdt;\de) & = & 
   {\cal F}_{\cpkl}^{\alpha,\rm sig}(\de) {\cal H}^{\alpha,\rm sig}_{tag,\cpkl}(\dt,\sdt) + \nn \\
  & & \sum_{j=\jpsi~X} {\cal F}^{\alpha,j}_{\cpkl} (\de)
     {\cal H}^{\alpha,j}_{tag,\cpkl}(\dt,\sdt) + \nn \\
& & {\cal F}^{\alpha,\rm non-\jpsi}_{\cpkl} (\de) \left[
    \sum_\beta f^{\alpha,\beta}_{\cpkl} {\cal H}^{\alpha,\beta}_{tag,\cpkl}(\dt,\sdt)
\right]~.\nonumber\\
\label{eq:pdfKl}
\eea
As the \jpsi\ lepton type is not expected to influence the \de\ shape,
the PDFs are used without regard to lepton type. The \de\ PDFs are
used separately for EMC and IFR \KL\ type, and they are grouped for
$\jpsi\KL$ (signal), $\jpsi\KS$ background, $\jpsi X$ background
(excluding $\jpsi\KS$) and non-\jpsi.

\subsection{Signal and background structure}

For signal events, a common set of mistag, \dt\ resolution function,
and $\nu$ and $\mu^{\alpha}$ parameters for all samples are assumed. This
assumption is supported by extensive Monte Carlo studies.  Peaking backgrounds originating 
in \Bz\ decays are assumed to have the same parameters as
the signal. For \Bu\ peaking backgrounds we still assume the same
resolution function as for signal, but the mistag and $\nu$, $\mu^{\alpha}$ 
parameters are fixed to the values extracted from the previous unbinned maximum-likelihood
fit to the \Bu\ data.  For combinatorial background components we use
an empirical description of the mistags and \dt\ resolution, allowing
various intrinsic time dependencies. In the nominal fit we assume 
prompt and non-prompt components, the non-prompt component being a
pure exponential dependence.  As discussed in
Sec.~\ref{sec:systematics}, fits with oscillatory, \CP/\T, \CPT/\CP\ and
\dckm\ structure have been also performed to evaluate possible
systematic biases.  A common set of mistags and \dt\ resolution
parameters, independent of the signal, is assumed for the non-\jpsi\
background in the \bcpkl\ sample and for the prompt and non-prompt
background components in the \bflav\ and \bcpks\ samples.  In this
case, the parameters $\dwa$ and $\wa_{\rm slope}$ are fixed to zero, and
the resolution model uses the core and outlier Gaussian distributions.
The fraction of prompt and non-prompt component and the exponential
constant of the non-prompt component in the non-\jpsi\ background are
fixed to the values obtained from an external fit to the time
distribution of the \jpsi\ dilepton mass sideband.  
The nominal fit neglects $\nu^{j}$ and $\mu^{\alpha,j}$ for combinatorial background sources.

\subsection{Free parameters for the nominal fit}

The ultimate aim of the fit is to obtain simultaneously
\sgndGoverG, \absqop, \reZparflat\ and \imZ, assuming $\Rcp=1$. 
The parameters \imlambcpflat\ and \dM\ are also floated to account for possible correlations and to provide an
additional cross-check of the measurements,
comparing our values with the \babar\ $\sintwob$ result based on the same data sample \cite{ref:sin2b-babar}
and recent \dM\ measurements.
The average \Bz\ lifetime, \taub,
is fixed to the PDG value, $1.542$ ps \cite{ref:pdg2002}. 
As a cross-check we also performed
fits with $R_{\CP}$ and $\taub$ allowed to vary.
All these physics parameters are by construction common to all samples and tagging
categories, although the statistical power for determining each
parameter comes from a particular combination of samples or \dt\
dependences, as discussed in Sec.~\ref{sec:decayrates}.

The doubly-\ckm-suppressed parameters are necessarily small and
difficult to determine.  In particular, \relambflavbare\ occurs only 
multiplied by other small parameters. As a result, we have
chosen to fix $\relambflavbare=0$.  We fit for the parameter
$ \imlambflavbare /|\lambflavbare|$, but fix $|A_{\rm flav}/\overline{A}_{\rm flav}|=0.02$.  
We regard these two variables as formal
entities, without the connection implied by taking $\relambflavbare=0$.
We vary separately $ \imlambbarflavbare /|\lambbarflavbare|$,
but keep $|\lambbarflavbare| = |\lambflavbare| |p/q|^2$.  
Thus there are two free
parameters, plus one fixed magnitude. We treat the tagging \dckm\
similarly.  We assign a systematic error by varying the magnitudes by
100\% and scanning all possible combinations of the phases
(Sec.~\ref{sec:systematics}). This strategy provides, for the current
data sample size, the optimal trade-off for all measured parameters
between statistical and systematic uncertainties originating in the
ambiguities of doubly-\ckm\ suppression in the \btag, and to a
lesser extent the \bflav.

The total number of parameters that are free in the fit is 58, of
which 36 parameterize the signal: physics parameters (4), cross-check
physics parameters (2), single effective imaginary parts of the
doubly-\ckm-suppressed phases (4), resolution function (10), mistags
(11) and differences in the fraction of \Bz\ and \Bzb\ mesons that are
tagged and reconstructed (5). 
The \dt\ distributions, the asymmetries and the physics parameters
\sgndGoverG, \absqop, \reZparflat, \imZ\ and the
cross-check parameter \imlambcpflat\ were kept hidden until the analysis
was finished. However, the parameter \dM, the residual \dt\ distributions and asymmetries,
the statistical errors and changes in the physics parameters due to
changes in the analysis were not hidden.

\section{\boldmath Analysis results}
\label{sec:results}

We extract the parameters \sgndGoverG, \absqop, \reZparflat\ and \imZ, \mbox{\imlambcpflat},
\dM, the parameters for doubly-\ckm-suppressed decays, 
the signal mistag, resolution function and $\nu$, $\mu^\alpha$ parameters and the empirical 
background parameters with the likelihood function
described in Sec.~\ref{sec:method}. In Table \ref{tab:signalyields} we list the signal yields
in each tagging category after vertexing requirements. The purities 
(estimated in the region $\mes>5.27$ \gevcc\ for non-\bcpkl\ samples 
and $| \Delta E | < 10$ \mev\ for \bcpkl\ events), averaged over
tagging categories, are 82\%, 94\% and 55\%, for \bflav, \bcpks\ and
\bcpkl\ candidates, respectively. The fitted signal mistag and
resolution function parameters are shown, respectively, in Tables
\ref{tab:mistag} and \ref{tab:resolution}. The values of the
asymmetries in reconstruction and tagging efficiencies are summarized
in Table \ref{tab:munu}. There is good agreement with the
asymmetries extracted with the counting-based approach outlined in
Appendix \ref{appendix:asymmetries}. Note that in the counting
technique the time-dependence of {\tt UnTagged} events is not used,
therefore decreasing significantly the statistical power of the
measurement of \dGoverG.

\begin{table}[htb!]
\begin{center}
\caption{Signal event yields, obtained from the \mes\ fits for the \bflav\ and \bcpks\ samples
and multiplying by the signal fraction in the $| \Delta E |<10$ \mev\ interval for the \bcpkl\ sample,
after vertexing requirements.\label{tab:signalyields}}
\vspace{0.1in}
\begin{tabular}{l|ccc|ccc|ccc} \hline
     &  \multicolumn{3}{|c|}{\bflav} & \multicolumn{3}{|c|}{\bcpks} & \multicolumn{3}{|c}{\bcpkl} \\
\hline
 Tag &   \Bz & \Bzb & Tot & \Bz & \Bzb & Tot & \Bz & \Bzb & Tot \\
\hline \hline
 {\tt Lepton}      & 1478 & 1419 & 2897 & 96  & 98  & 194 & 35 & 35 & 70 \\
 {\tt Kaon I}    & 2665 & 2672 & 5337 & 154 & 175 & 329 & 74 & 65 & 139 \\
 {\tt Kaon II} & 3183 & 2976 & 6159 & 181 & 188 & 369 & 85 & 66 & 151 \\
 {\tt Inclusive}       & 3197 & 3014 & 6211 & 184 & 172 & 356 & 78 & 72 & 150 \\
 {\tt UnTagged}    & \multicolumn{3}{|c|}{10423} & \multicolumn{3}{|c|}{585} & \multicolumn{3}{|c}{260} \\
\hline
\end{tabular}
\end{center}
\end{table}

\begin{table}[!htb] 
\begin{center}
\caption{Signal tagging efficiencies and mistag parameters for each
  tagging category $\alpha$ as extracted from the nominal
  maximum-likelihood fit. Uncertainties are statistical only.}
\vspace{0.1in}
\label{tab:mistag} 
\begin{tabular}{l|c|c|c|c}
\hline
Tagging category   & $t^\alpha(\%)$ & $\wa_0$ & $\wa_{\rm slope}$ & $\dwa$ \\
\hline \hline
{\tt Lepton}    & $9.4  \pm  0.2$ & $0.026\pm0.007$ & 0 (fixed) & $-0.012 \pm 0.012$  \\
{\tt Kaon I}    & $17.2 \pm  0.3$ & $0.020\pm0.020$ & $0.13\pm0.04$ & $-0.027 \pm 0.013$  \\
{\tt Kaon II}   & $19.9 \pm  0.3$ & $0.159\pm0.024$ & $0.07\pm0.04$ & $-0.042 \pm 0.013$  \\
{\tt Inclusive} & $19.9 \pm  0.3$ & $0.265\pm0.025$ & $0.07\pm0.04$ & $-0.029 \pm 0.013$  \\
{\tt UnTagged}  & $33.6 \pm  0.6$ & --- & ---  & ---  \\
\hline
\end{tabular} 

\end{center}
\end{table}

\begin{table}[!htb] 
\begin{center}
\caption{Signal \deltat\ resolution function parameters as extracted
  from the nominal maximum-likelihood fit. Uncertainties are
  statistical only.}
\vspace{0.1in}
\label{tab:resolution} 
\begin{tabular}{l|c||c|c}
\hline
Parameter & Fitted value & Parameter & Fitted value\\ 
\hline \hline
$S_{\rm core}$                          & $1.25 \pm 0.04$ & 
$S_{\rm tail}$                          & $5.7\pm0.8$  \\
$\delta_{\rm core}^{\tt Lepton}$        & $0.02 \pm 0.07$ &
$\delta_{\rm tail}$                     & $-1.5 \pm 0.5$ \\
$\delta_{\rm core}^{\tt Kaon I}$        & $-0.27 \pm 0.05$ &
$f_{\rm tail}$                          & $0.034 \pm 0.010$ \\
$\delta_{\rm core}^{\tt Kaon II}$       & $-0.32 \pm 0.04$ &
$\sigma_{\rm out}$                      & 8~ps (fixed)  \\
$\delta_{\rm core}^{\tt Inclusive}$     & $-0.30 \pm 0.04$ &
$\delta_{\rm out}$                      & 0~ps (fixed) \\
$\delta_{\rm core}^{\tt UnTagged}$      & $-0.28 \pm 0.03$  &  
$f_{\rm out}$                           & $0.0003 \pm 0.0012$ \\
\hline
\end{tabular} 

\end{center}
\end{table}

\begin{table}[!htb] 
\begin{center}
\caption{Values of the signal $\Bz\Bzb$ differences in reconstruction ($\nu$) and 
tagging ($\mu^{\alpha}$) efficiencies as extracted from the nominal maximum-likelihood fit. 
The results are compared with those obtained with a counting-based method.}\
\vspace{0.1in}
\label{tab:munu} 
\begin{tabular}{l|c|c}
\hline
Parameter       & Nominal fit & Counting-based method \\
\hline\hline
$\nu$                 & $ 0.011 \pm 0.008$ & $ 0.007\pm0.008$ \\
$\mu^{\tt Lepton}$    & $ 0.024 \pm 0.022$ & $ 0.029\pm0.042$ \\
$\mu^{\tt Kaon I}$    & $-0.022 \pm 0.017$ & $-0.022\pm0.029$ \\
$\mu^{\tt Kaon II}$   & $ 0.014 \pm 0.016$ & $ 0.004\pm0.027$ \\
$\mu^{\tt Inclusive}$ & $ 0.014 \pm 0.016$ & $ 0.025\pm0.027$ \\
\hline
\end{tabular} 

\end{center}
\end{table}


The values of \sgndGoverG, \absqop, \reZparflat\ and \imZ\ extracted
from the fits are given in Table \ref{tab:phys-param}.  
The fitted effective doubly-\ckm-suppressed decay parameters are also
indicated. All these results can be compared to those obtained when
the fit was repeated assuming \CPT\ invariance. The significant change
of the effective doubly-\ckm-suppressed decay parameters between
the two fits is due to the large correlation of these parameters with
the \CPT\ violating parameter \imZ. The fitted value of \dM\ agrees with
recent \B-Factory measurements \cite{ref:dM-babar-had,ref:dM-babar-dstlnu,ref:dM-babar-dilep,ref:dM-belle}, 
and remains unchanged between the two fits. The fit result for \imlambcpflat\ when we assume \CPT\
invariance 
agrees with our $\sintwob$ measurement
based on the same data set \cite{ref:sin2b-babar}. When we allow for
\CPT\ violation, \imlambcpflat\ increases by $+0.012$ with unchanged statistical
errors.  The statistical correlation coefficients among all
physics and cross-check physics parameters
are shown in Table \ref{tab:correlation}. Table \ref{tab:correlation-top} shows 
the top five statistical correlations of the physics parameters with any other 
free parameter in the global fit.

\begin{table}[!htb] 
\begin{center}
\caption{Physics parameters results, from the global nominal fit and when we assume \CPT\ invariance.
The free single effective doubly-\ckm-suppressed decay parameters are also indicated.
Errors are statistical only.} 
\label{tab:phys-param} 
\vspace{0.1in}
\begin{tabular}{l|c|c}
\hline
Parameter &  Nominal fit results & Fit results assuming \CPT\ invariance \\ 
\hline \hline 
$\sgndGoverGd$       &  $ -0.008 \pm 0.037 $ & $ -0.009 \pm 0.037 $     \\
$\absqop$         &  $ 1.029 \pm 0.013 $ & $ 1.029 \pm 0.013 $      \\
$\reZparflat$  &  $ 0.014 \pm 0.035 $ & $-$ \\
$\imZ$            &  $ 0.038 \pm 0.029 $ & $-$ \\
\hline 
$\imlambtagflat$      &  $ 1.5 \pm 1.2 $ & $ 0.5 \pm 1.0 $ \\
$\imlambbartagflat$   &  $ -0.1 \pm 1.2 $ & $ 0.8 \pm 1.0 $ \\
$\imlambflavflat$     &  $ 2.3 \pm 1.1 $ & $ 1.4 \pm 0.9 $ \\
$\imlambbarflavflat $ &  $ -0.6 \pm 1.1 $ & $ 0.1 \pm 0.9 $ \\
\hline
\end{tabular}

\end{center}
\end{table}

\begin{table}[!htb] 
\begin{center}
\caption{Correlation among all the physics parameters as extracted from the 
simultaneous maximum-likelihood fit to the \bflav\ and \bcp\
samples. 
}
\label{tab:correlation}  
\vspace{0.1in}
\begin{tabular}{l||c|c|c|c|c} 
\hline
& $\sgndGoverG$ & \absqop & \imlambcpflat & \reZparflat & \imZ\\ 
 \hline \hline 
$ \dM$  & $  -1.3 \% $ & $  -2.8 \% $ & $  -5.6 \% $ & $  7.0 \% $ & $  -0.2 \% $\\ 
$ \sgndGoverGd$  &  & $  11.0 \% $ & $  0.4 \% $ & $  -7.9 \% $ & $  -1.8 \% $\\ 
$ \absqop$ & & & $  -1.0 \% $ & $  -2.4 \% $ & $  -1.1 \% $\\ 
$ \imlambcpflat$  & & & & $  -10.9 \% $ & $  17.4 \% $\\ 
$ \reZparflat$  &  &  &  &  & $  -3.4 \% $\\ 
\hline
\end{tabular}

\end{center}
\end{table} 

\begin{table}[!htb] 
\begin{center}
\caption{Top five correlations of the physics parameters with any other free parameter
of the global maximum-likelihood fit.}
\label{tab:correlation-top}  
\vspace{0.1in}
\begin{tabular}{l|l|r} 
\hline
Physics parameter & Parameter  & Correlation (\%) \\
\hline \hline
\dM  & $S_{\rm tail}$ & $-15.4$ \\
     & \reZparflat & $-5.8$ \\
     & $w^{\tt KaonI}_0$ & $-5.8$ \\
     & $w^{\tt KaonII}_{\rm slope}$ & $-4.7$ \\
     & $\delta_{\rm core}^{\tt UnTagged}$ &  $4.2$ \\
\hline
\sgndGoverG & \absqop & $11.0$ \\
         & \reZparflat &  $-7.9$ \\
         & $\nu$ & $6.8$ \\
         & $\dw^{\tt KaonI}$ & $-3.8$  \\
         & $\dw^{\tt KaonI}$ & $-3.4$ \\
\hline
\absqop & $\nu$ & $65.1$ \\
        & $\dw^{\tt KaonII}$ & $-22.5$ \\
        & $\dw^{\tt KaonI}$ & $-22.4$ \\
        & $\dw^{\tt Inclusive}$ & $-15.5$  \\
        & $\mu^{{\tt KaonII}}$ & $3.2$ \\
\hline
\imlambcpflat & \imZ & $17.4$\\
          & \imlambtagflat & $14.4$ \\
          & \imlambbarflavflat & $-6.2$ \\
          & \imlambbartagflat & $-5.5$ \\
          & $w^{\tt KaonII}_0$ & $4.5$ \\
\hline
\reZparflat    & \imlambcpflat & $-10.9$ \\
               & \imZ & $-3.4$ \\
               & $\mu^{{\tt Lepton}}$ & $2.2$ \\
               & \imlambtagflat & $-2.0$ \\
               & $w^{\tt KaonII}_0$ &  $-1.6$ \\
\hline
\imZ & \imlambtagflat & $61.6$ \\
     & \imlambbartagflat & $-56.6$  \\
     & $\dw^{\tt KaonI}$ &  $-8.0$ \\
     & $\dw^{\tt KaonII}$ &  $-6.1$ \\
     & $\dw^{\tt Inclusive}$ &  $-3.5$ \\
\hline
\end{tabular}

\end{center}
\end{table}

\begin{figure}[h!]
\begin{center}   
\begin{tabular} {c}  
\epsfig{file=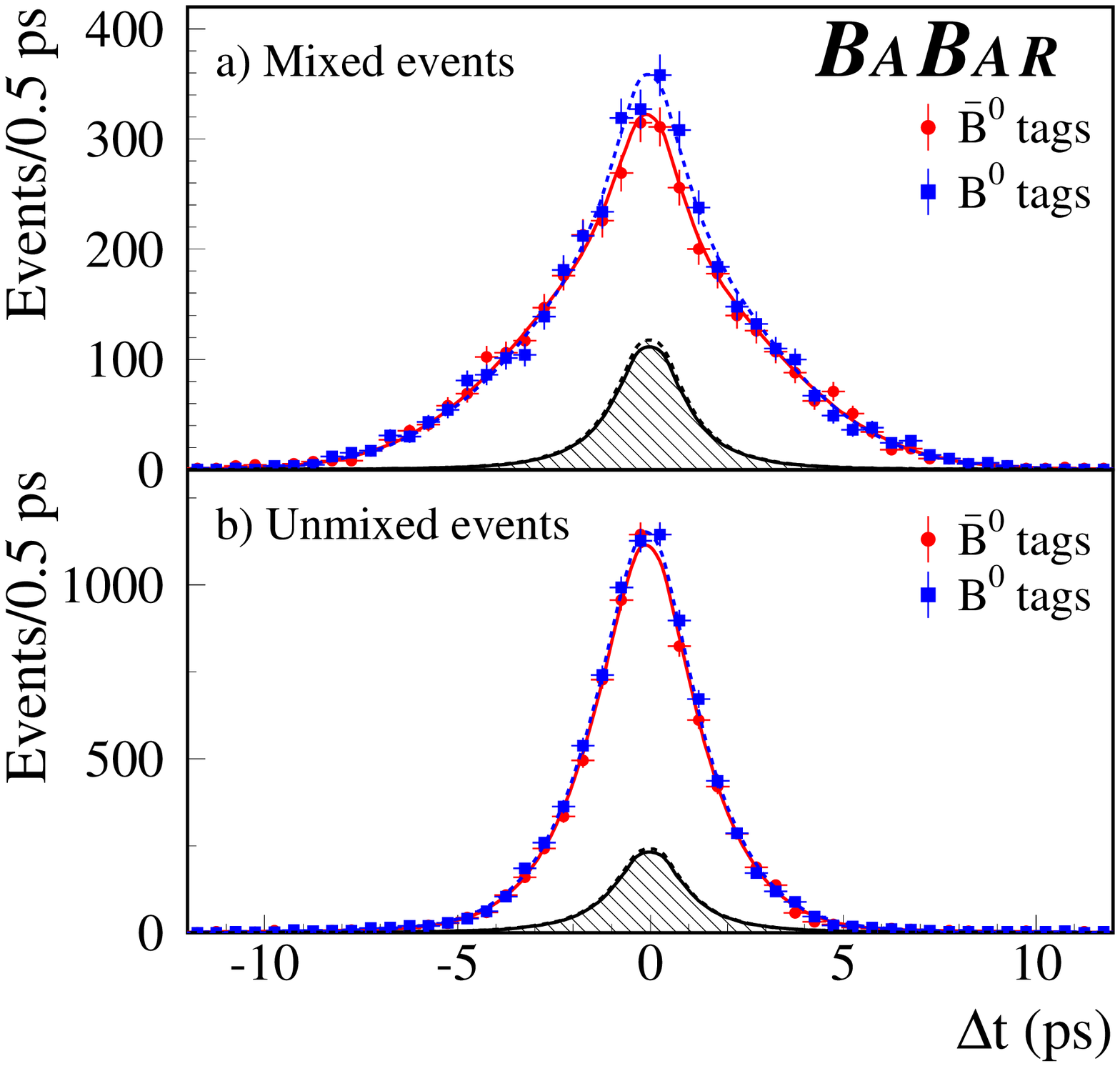,width=0.7\linewidth} 
\end{tabular} 
\end{center} 
\caption{The \dt\ distributions for (a) \bflav\ mixed and (b) \bflav\ unmixed events with a \Bz\ tag and with a \Bzb\ tag
in the signal region, $\mes>5.27$ \gevcc. The solid (dashed) curves represent
the fit projection in \dt\ based on the individual signal probabilities and event-by-event
\dt\ uncertainty for \Bzb(\Bz) tags. 
The shaded area shows the background contribution to the distributions. 
\label{fig:dtbflav}}
\end{figure}

\begin{figure}[h!]
\begin{center}   
\begin{tabular} {c}  
\epsfig{file=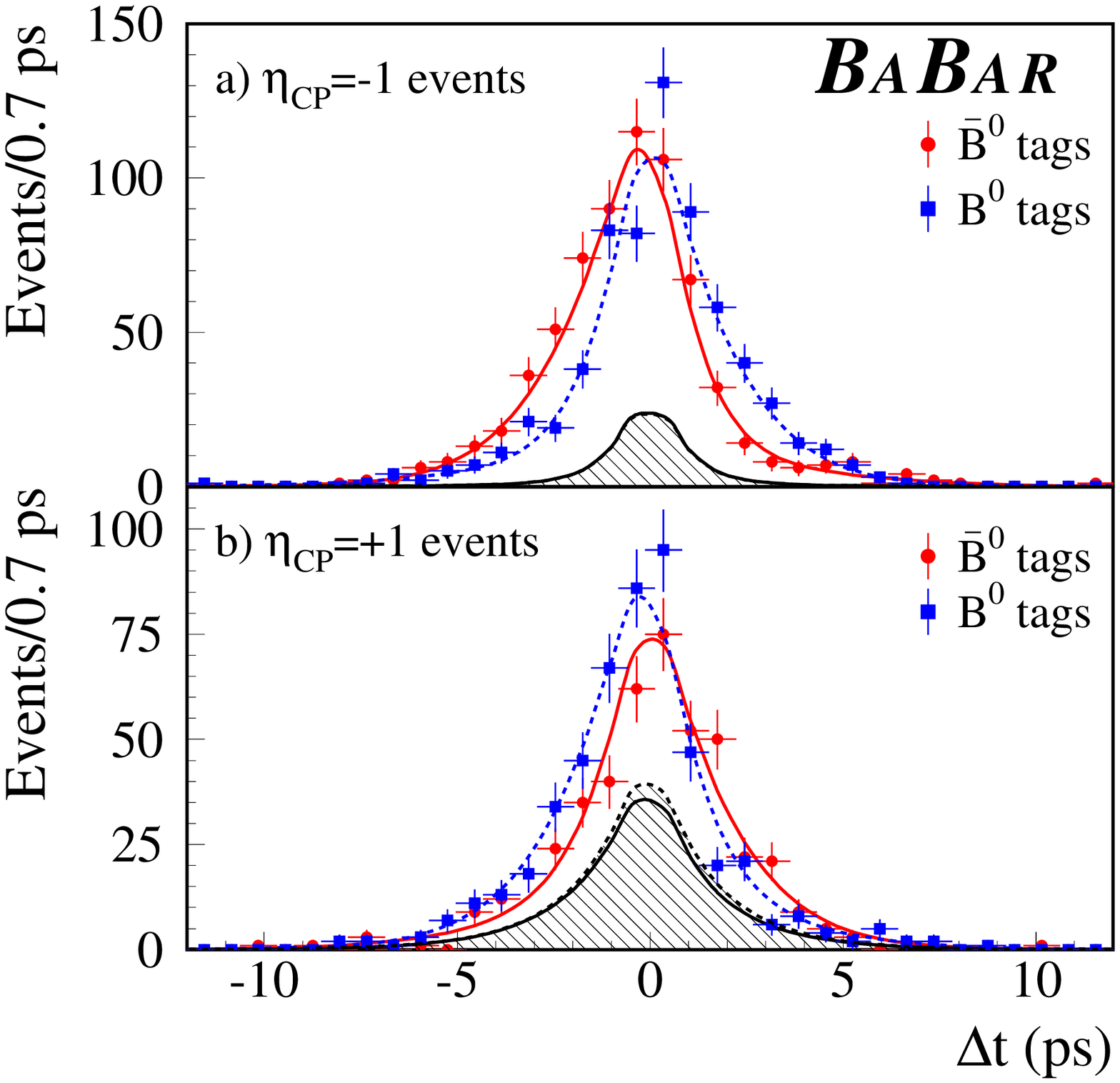,width=0.7\linewidth}
\end{tabular}   
\end{center} 
\vspace{-0.5in}
\caption{The \dt\ distributions for (a) \bcpks\ and (b) \bcpkl\ events with a \Bz\ tag and with a \Bzb\ tag
in the signal region, $\mes>5.27$ \gevcc\ for \bcpks\ candidates 
and $| \Delta E | < 10$ \mev\ for \bcpkl\ events. The solid (dashed) curves represent
the fit projection in \dt\ based on the individual signal probabilities and event-by-event
\dt\ uncertainty for \Bzb(\Bz) tags. 
The shaded area shows the background contribution to the distributions. 
\label{fig:dtbcp}}
\end{figure}

Figures \ref{fig:dtbflav} and \ref{fig:dtbcp} show the \dt\
distributions of the signal candidates ($\mes>5.27$ \gevcc\ for
\bflav\ and \bcpks\ and $| \Delta E | < 10$ \mev\ for \bcpkl\ samples). The
points correspond to data. The curves 
correspond to the projections of the nominal likelihood fit
weighted by the appropriate relative amounts of signal and
background. The background contribution is indicated with the shaded area. 


\section{\boldmath Cross-checks and validation studies}
\label{sec:checks}

We have used data and Monte Carlo samples to perform validation
studies of the analysis technique.  Monte Carlo tests include studies
with parameterized fast Monte Carlo as well as full 
GEANT4 \cite{ref:geant4} simulation samples. Checks with data were performed
with control samples where no \dG, \CP/\T or \CPT/\CP\ effects are expected.  Other
checks were made by analyzing the actual data sample, but using
alternative tagging, vertexing and fitting configurations.

\subsection{Monte Carlo simulation studies}
\label{sec:checks-mc}

An extensive test of the fitting procedure was performed with fast
parameterized Monte Carlo simulations, generating 300 experiments
 with a sample size and composition corresponding to that
of the data. The mistag
rates and \dt\ distributions were generated according to the model used
in the likelihood function. The nominal fit was then performed on each
of these experiments.  Each experiment used the set of \mes($\Delta
E$) and $\sdt$ values observed in the non-\KL(\KL) sample.  The
rms spread of the residual distributions for all physics parameters
(where the residual is defined as the difference between the fitted
and generated values), was found to be consistent, within 10\%, with
the mean (Gaussian) statistical errors reported by the fits. Moreover,
from these experiments it was verified that the asymmetric 68\% and
90\% confidence intervals provide the correct coverage.  The mean
values of the residual distributions were in all cases consistent with
no measurement bias. A systematic error due to the precision of this
study was assigned to each physics parameter.  The statistical errors
on all the physics parameters (Table \ref{tab:phys-param}) and the
calculated correlation coefficients among them (Tables
\ref{tab:correlation} and \ref{tab:correlation-top}) extracted from the nominal fit to the data were
consistent with the range of values obtained from these experiments: We
found that 24\% of the fits result in a value of the negative
log-likelihood that is less (better) than that found in data.

In addition, large samples of signal and background Monte Carlo events
generated with a 
full
detector simulation were used
to validate the measurement. The largest samples were generated with
zero values of \dGoverGd, $\absqop-1$ and $z$, but additional samples
were also produced with relatively large values of these parameters. Other values
(including those measured in the data) were generated with reweighting
techniques. The signal Monte Carlo samples were split into
samples comparable to the actual data set, keeping the relative sizes of signal \bflav,
\bcpks\ and \bcpkl\ samples as observed. To check whether
the selection criteria, or the analysis and fitting procedure,
introduced any bias in the measurements, the nominal fit (signal only)
was then applied to these experiments. The small combinatorial
background in these signal samples was rejected, using only events in
the signal region.  Fits to the pure signal physics model, using the
true \dt\ distribution and true tagging information, were also
performed.  The means of the residual distributions from all these
experiments for all the physics parameters are consistent with zero,
confirming that there is no measurement bias. The rms spreads are
consistent with the average reported errors and with the estimated
errors in the fit to data.  A systematic error is assigned to each
physics parameter due to the limited Monte Carlo statistics from this
test. The effect of backgrounds has been evaluated by adding an
appropriate fraction of background events to the signal Monte Carlo
sample and performing the fit. The \bcp\ background samples were
obtained either from simulated $\B \to \jpsi X$ events or $\Delta E$
sidebands in data, while the \bflav\ backgrounds were obtained from
generic $\B\Bb$ Monte Carlo. We find no evidence of bias in any of the
physics parameters.

\subsection{Cross-checks with data}
\label{sec:checks-data}

We fit subsamples selected by choosing a
tagging category and running period. Fits using only the 
$\Bz \to D^{(*)-} X^+$ or $\Bz \to \jpsi \Kstarz(\Kp\pim)$ specific
channels for \bflav, and \bcpks\ or \bcpkl\ only for \bcp, were also
performed. We found no statistically significant differences in the results
for the different subsets.  We also varied the maximum values of $| \dt |$ 
and $\sdt$ accepted between 5 and 30 ps, and 0.6 and 2.2 ps,
respectively.  Again, we did not find statistically significant changes
in the fitted values of the physics parameters.

In order to verify that the results are stable under variation of the
vertexing algorithm that is used in the measurement of \dt, we used
alternative (less powerful) methods \cite{ref:babar-stwob-prd}. In
order to reduce statistical fluctuations due to different events being
selected, the comparison between the alternative and nominal methods
was performed using only the common events. Observed variations were consistent
with anticipated statistical fluctuations and with the systematic error assigned to the
resolution function (see Sec.~\ref{sec:systematics}).

The stability of the results under variation of the tagging algorithm
was studied by repeating the fit using the tagging algorithm used in
Ref.~\cite{ref:babar-stwob-prd}. The algorithm used in that analysis
had an effective tagging efficiency, $Q=\sum_{\alpha}
t^{\alpha}\left(1-2\wa \right)^2$, about 7\% lower than the one used here. 
The variation observed in the physics parameters is consistent with
the statistical differences.

The average \Bz\ lifetime was fixed in the nominal fit to the PDG value
\cite{ref:pdg2002}. This value was obtained by averaging
measurements based on flavor eigenstate samples and assuming
negligible effects from \dGoverGd, \absqop\ and \CPT\ violation. 
Measurements that do not use tagged events are not
affected by \absqop\ and \CPT\ violation, but are by a non-zero value
\dGoverGd\ at second order, as discussed 
in Sec.~\ref{sec:decayrates}. Therefore we do not expect significant
effects from the fixed average \Bz\ lifetime.  However, to check the
consistency of the result, the fit was repeated with \taub\
left free. The resulting \taub\ was about two standard deviations
below the nominal value assumed in our analysis, 
before bias corrections and taking into account the statistical
error from the fit and the present \taub\ uncertainty. 
The changes of the physics parameters were within
statistical errors. Nevertheless, a systematic
error was assigned using the variation of each physics parameter
repeating the fit with \taub\ fixed after a change of two standard
deviations of the present uncertainty ($\pm 0.032$ ps).

Similarly, fits with \Rcp\ floated were performed. The resulting \Rcp\
value was consistent with 1 (the fixed nominal value), within one
standard deviation (statistical only).  The changes observed in the
physics parameters were consistent with their statistical uncertainties.
Systematic errors due to fixing \Rcp\ at unity were set by changing
\Rcp\ by twice the statistical uncertainty determined by floating it
($\pm 10\%$).  The resulting variation in each parameter was taken as
the systematic error.

The robustness of the fit was also tested by modifying the nominal PDF
normalization, as described by Eq. (\ref{eq:norm}), so that the
analysis was insensitive to the relative amount of \Bz\ and \Bzb\
tagged events. As a consequence, the statistical power of \absqop\ was
dramatically reduced, since the sensitivity of this parameter comes
largely from the differences in time-integrated \Bz\ and \Bzb\ rates.
In addition, the fit was also performed assuming an independent set of
resolution function parameters for each tagging category. In all cases
the results are compatible within statistical differences with the
nominal fit results. Finally, the tagging efficiencies $t^{\alpha,j}$
were alternatively determined from the \bflav\ sample and assumed to
be the same for all samples (rather than to use the estimate from each
sample separately, as in the nominal fit).  The change in the values
of the physics parameters was found to be negligible.

Control samples in data from \Bu\ decays (treated in a fashion analogous 
to that described in 
Sec.~\ref{sec:sample}) have also been used to validate the analysis
technique, since in these samples we expect zero values for \dGoverG,
\absqop\ and $z$. For the \bflav\ sample we used the $\Bu \to
\overline{D}^{(*)0}\pip$ decay channel, and for the \bcp\ sample the
charmonium \Bu\ decays.  Due to the absence of mixing and \CP\ violation
in these samples, the check was performed fixing $\dM=0$ and
$\absqop=1$ in the \bflav\ sample, and $\dM=0.489$ ps$^{-1}$ and
$\imlambcpflat=0$ in the \bcp\ sample, fitting only for \sgndGoverGd,
\reZparflat\ and \imZ.  No statistically significant deviations from
zero were observed.

\section{\boldmath Systematic uncertainties}
\label{sec:systematics}

We estimate systematic uncertainties with
studies performed on both data and Monte Carlo simulation samples. A
summary of the non-negligible sources and results is shown in Table
\ref{tab:systglob}.  In the following, the individual 
contributions are referenced by the lettered lines in this table.

\begin{table}[h!]
\caption{Summary of systematic uncertainties on the measurement of \sgndGoverG, 
\absqop, \reZparflat\ and \imZ.}
\begin{center}
\renewcommand{\arraystretch}{1.3}
\begin{tabular}{l|c|c|c|c}
\hline
Systematics source & \sgndGoverG & \absqop & \reZparflat & \imZ \\ 
\hline\hline
\multicolumn{5}{c}{Likelihood fit procedure} \\
\hline
(a) Parameterized MC test & 
$  0.003 $ & $  0.001 $ & $  0.003 $ & $  0.003 $ \\
(b) GEANT4 MC test &
$  0.005 $ & $  0.007 $ & $  0.004 $ & $  0.016 $ \\
\hline \hline
\multicolumn{5}{c}{\dt\ resolution function} \\
\hline
(c) Res. funct. parameterization  &
$  0.007 $ & $  0.001 $ & $  0.008 $ & $  0.003 $ \\
(d) $z$ scale and boost  &
$  0.003 $ & $  0.001 $ & $  0.002 $ & $ <0.001 $ \\
(e) Beam spot  & 
$  0.008 $ & $  0.002 $ & $  0.001 $ & $  0.011 $ \\
(f) SVT alignment  &
$  0.006 $ & $  0.001 $ & $  0.001 $ & $  0.011 $ \\
(g) Outliers   &
$  0.002 $ & $ <0.001 $ & $ <0.001 $ & $ <0.001 $ \\
\hline \hline
\multicolumn{5}{c}{Signal properties} \\
\hline
(h) Average $\Bz$ lifetime  &
$  0.004 $ & $  0.001 $ & $  0.004 $ & $ <0.001 $ \\
(i) Direct \CP\ violation  &
$  0.002 $ & $  0.004 $ & $  0.001 $ & $  0.003  $ \\
(j) Doubly-\ckm-suppressed decays  &
$  0.008 $ & $  0.004 $ & $  0.032 $ & $  0.006  $ \\
(k) Residual charge asymmetries  &
$  0.005 $ & $  0.006 $ & $  0.004 $ & $  0.006 $ \\
\hline \hline
\multicolumn{5}{c}{Background properties and structure} \\
\hline
(l) Signal probability  & 
$  0.002 $ & $  0.001 $ & $  0.002 $ & $  0.001 $ \\
(m) Fraction of peaking background  &
$ <0.001 $ & $ <0.001 $ & $  0.004 $ & $ <0.001 $ \\
(n) $\dt$ structure  &
$  0.002 $ & $  0.001 $ & $  0.001 $ & $  0.001 $ \\
(o) \dGd/\CPT/\CP/\T/Mixing/\dckm\ &
$  0.001 $ & $  0.002 $ & $  0.002 $ & $ <0.001 $ \\ 
(p) Residual charge asymmetry  &
$ <0.001 $ & $  0.001 $ & $ <0.001 $ & $ <0.001 $ \\ 
(q) \kl\ specific systematics &
$  0.004 $ & $ <0.001 $ & $  0.004 $ & $  0.003 $ \\
\hline\hline
Total systematics & {\bf 0.018}    & {\bf 0.011} & {\bf 0.034} & {\bf 0.025}\\ 
\hline
\end{tabular}
\renewcommand{\arraystretch}{1.0}


\end{center}
\label{tab:systglob}
\end{table}

\subsection{Likelihood fit procedure}

Several sources of systematics due to the likelihood fit procedure
are considered. We first include the results from the tests
performed using the fast parameterized Monte Carlo (a) and the full
GEANT4 signal Monte Carlo events (b), as described in
Sec.~\ref{sec:checks-mc}.  We take the larger of the observed bias (mean of the residual distributions)
and its statistical error as the systematic error.  No corrections are applied to the central
values extracted from the fit to the data since no statistically
significant bias is observed. 
Note that the GEANT4 simulation addresses the underlying assumption
that the fit properly accounts for residual differences in the mistag,
resolution function, $\nu$, $\mu^\alpha$ parameters for \bflav,
\bcpks\ and \bcpkl\ samples and differences in \dt\ resolution for
correct and wrong tags \cite{ref:babar-stwob-prd}.

We also consider the impact on the measured physics parameters of the
asymptotic PDF normalization. The effect is evaluated by varying the
fitted values using a normalization in the range defined by the \dt\ cut. Finally,
the fixed tagging efficiencies are varied within their statistical uncertainties.
The two contributions are found to be negligible.

\subsection{\dt\ resolution function}

The resolution model used in the analysis, consisting of the sum of
three Gaussian distributions, is expected to be flexible enough to
accommodate the actual resolution function. To assign a systematic
error to the assumption of this model, we use the alternative model
described in Sec.~\ref{sec:decaytime}, with a Gaussian distribution
plus the same Gaussian convolved with one exponential function, for
both signal and background.  The results for all physics parameters
obtained from the two resolution models are consistent within
statistical uncertainties. However, we assign the difference of
central values as a systematic uncertainty (c).

In addition a number of parameters that are integral to the
determination of \dt\ are varied according to known uncertainties. The
PEP-II boost, estimated from the beam energies, has an uncertainty of
0.1\% \cite{ref:babar-nim}. The absolute $z$-scale uncertainty has
been evaluated to be less than 0.4\%. This estimate was obtained by
measuring the beam pipe dimensions with scattered protons and
comparing to optical survey data. Therefore, the boost and $z$-scale
systematics are evaluated by varying by $\pm0.6\%$ the reconstructed
\dt\ and $\sdt$ (d).  The uncertainty on the beam spot, which is much
wider than it is high, is taken into account by moving its vertical
position (the direction most valuable in vertexing) by 20 and 40
$\mu$m and increasing the vertical dimension by 30 and 60 $\mu$m
(e). Finally, the systematic uncertainty due to possible SVT internal
misalignment is evaluated by applying a number of possible misalignment
scenarios to a sample of simulated events and comparing the values of
the fitted physics parameters from these samples to the case of
perfect alignment (f).

Fixing the width and bias of the outlier component, respectively to 8.0 and 0.0 ps,
introduces systematic errors.
To estimate the uncertainty we add in quadrature the
variation observed in the physics parameters when the bias changes
by $\pm$5 ps, the width varies between 6 and 12 ps and the outlier
distribution is assumed to be flat (g).

\subsection{Signal properties}

As described in Sec.~\ref{sec:checks-data}, the uncertainty from fixing the 
average \Bz\ lifetime has been evaluated by moving its central value by  $\pm 0.032$ ps (h), 
twice the current 
uncertainty \cite{ref:pdg2002}. Possible direct \CP\ violation in the \bcp\ sample
was taken into account by varying \Rcp\ by $\pm 10\%$ (i).

Systematics from doubly-\ckm-suppressed decays arise due to uncertainties
in \relambtagbare, \relambbartagbare, \relambflavbare\ and \relambbarflavbare.
In order to evaluate this contribution, samples of parameterized Monte
Carlo tuned to the data sample, with all possible values of the
doubly-\ckm-suppressed phases were generated, assuming a single
hadronic decay channel contributing to the \btag\ and to the \bflav.
The generation was made for maximal values of $| A_{\rm tag} / \overline{A}_{\rm tag} |$
and $| A_{\rm flav} / \overline{A}_{\rm flav} |$, assuming a 100\% uncertainty on its
estimate based on the elements of the CKM matrix, $\approx 0.02$
\cite{ref:pdg2002}.  For the {\tt Lepton} tagging category, largely
dominated by semileptonic \B\ decays, we assumed $| \lambtagbare |$ to be zero. 
Using the fit results from all these samples, we
evaluated the larger of the offset with respect to the generated
value and its statistical uncertainty, for each possible configuration
of phases. The systematic error assigned was  the largest value
among all configurations (j).  This is the dominant source of
systematic uncertainty for the measurement of \reZparflat, primarily
due to the \dckm\ effects in the tagging \B\ meson. Similar studies performed
with more than one hadronic channel indicated the destructive
interference among the different decay modes, proving that our
prescription to assign the systematics assuming a single effective
channel is conservative. 

Charge asymmetries induced by a difference in the detector response
for positive and negative tracks are included in the PDF and extracted
together with the other parameters from the time-dependent
analysis. Thus, they do not contribute to the systematic error, but
rather are incorporated into the statistical error at a level
determined by the size of the \bflav\ data sample. Nevertheless, in
order to account for any possible and residual effect, we assigned a
systematic uncertainty as follows.  We reran the \B\ reconstruction,
vertexing and tagging code after killing randomly and uniformly (no
momentum or angular dependence) 5\% of positive and negative
tracks in the 
full Monte Carlo sample.
This 5\% is on average more than a factor three larger than the
precision with which the parameters $\nu$ and $\mu^{\alpha}$ have been
measured in the data.  The half difference between the results
obtained for positive and negative tracks is assigned as a systematic error
(k).

\subsection{Background properties and structure}

The event-by-event signal probability $p^{\alpha,\rm
sig}_{rec}(\mes)$ for \bflav\ and \bcpks\ samples was fixed to the
values obtained from the \mes\ fits. We compared the results from
the nominal fits to the values obtained by changing one sigma up and
down all the \mes\ distribution parameters, taking into account their
correlations. This was performed simultaneously for all tagging
categories, and independently for the \bflav\ and \bcpks\
samples. Alternatively, we also used a flat signal probability
distribution: events belonging to the sideband region (\mes$<$5.27
\gevcc) are assigned a signal probability of zero, while we gave a
signal probability equal to the purity of the corresponding sample to
signal region events (\mes$>$5.27 \gevcc). The differences among
fitted physical parameters with respect to the default method were
found to be consistent within the statistical differences. 
We determined the systematic error due to this parameterization by varying the 
signal probability by the statistical error in the purity. The final
systematic error was taken to be the larger of the one-sigma variations
found for the two methods (l). The uncertainty on the fraction of
peaking background was estimated by varying the fractions according to
its uncertainty separately for the \bflav\ sample and each \bcpks\
decay mode (m). The effective $\eta_{\CP}$ of the
\bcpks\ peaking background, assumed to be zero 
in the nominal fit, was also varied between $+1$ and $-1$
and found to be negligible.

Another source of systematic uncertainty originates from the
assumption that the \dt\ structure of the combinatorial background in
the \mes\ sideband region is a good description of the structure in the signal
region. However, the background composition changes slightly as a function of \mes, since the
fraction due to continuum production slowly decreases towards the \B\ mass. To study 
this effect, we first varied the lower edge of the \mes\ distributions 
from 5.20 \gevcc\ to 5.27 \gevcc, simultaneously for the
\bflav\ and \bcpks\ samples, observing a good stability of the result. 
We also split the sideband region in seven equal slices 
each 10 \mevcc\ wide and used each of these ranges to perform a standard fit.
The quadratic sum of the extrapolation to the \B\ mass region and the error on it
was assigned as systematic uncertainty (n).

As described in Sec.~\ref{sec:method}, the nominal likelihood fit
assumes that there is no \dG, \CP/\T, \CPT/\CP, mixing and
doubly-\ckm-suppressed decays content in the combinatorial background
components (\bflav\ and \bcpks\ samples) and in the 
non-\jpsi\ background (\bcpkl\ sample). To evaluate the effect of this assumption
we repeated the fit but now assuming non-zero values of \dG, \absqop,
$z$, \imlambcpbare\ and \dM, varying $\eta_{\CP}$ of the background by $\pm
1$.  The check was performed by introducing in the PDF an independent
set of physics parameters and assuming maximal mixing and \CP\ violation
(\dM\ and \imlambcpflat\ fixed to 0.489 ps$^{-1}$ \cite{ref:pdg2002} and
0.75 \cite{ref:sin2b-babar}, respectively). Doubly-\ckm-suppressed
decay effects were included assuming the maximal values of
$| A_{\rm tag} / \overline{A}_{\rm tag} |$, $| A_{\rm flav} / \overline{A}_{\rm flav} |$ and
scanning all the possible values of the \Bz\ and \Bzb\ phases for \bflav\ and \btag. The systematic
uncertainty was evaluated simultaneously for all of these sources (o).

The uncertainty due to the \Bu\ lifetime has been evaluated by moving
the central value according to the current uncertainty
\cite{ref:pdg2002}. It was found to be negligible. The \Bu\ mistags
and the differences in the fraction of \Bu\ and \Bub\ mesons that are
tagged and reconstructed were varied according to their statistical
errors as obtained from the fit to the \Bu\ data. They were found also
to be negligible.  Uncertainties from charge asymmetries in
combinatorial background components (neglected in the nominal fit)
were evaluated by repeating the fit with a new set of $\nu$ and
$\mu^\alpha$ parameters. The measured values of $\nu$ and $\mu^\alpha$
are found to be compatible with zero and the variation of the physical
parameters with respect to the nominal fit is assigned as systematic
error (p).

For the \bcpkl\ channel \cite{ref:babar-stwob-prd}, the signal and
non-\jpsi\ background fractions are varied according to their
statistical uncertainties as obtained from the fit to the \de\
distribution.  We also vary background parameters, including the
$\jpsi X$ branching fractions, the assumed $\eta_{\CP}$, the \de\ shape
and the fraction and effective lifetime of the prompt and non-prompt
non-\jpsi\ components. The differences observed between data and Monte
Carlo simulation for the \KL\ angular resolution and for the fractions
 of $\Bz \to \jpsi\KL$ events reconstructed in the EMC and IFR
are used to evaluate a systematic uncertainty due to the simulation of the \KL\
reconstruction.
Finally, an additional contribution is assigned to the correction
applied to {\tt Lepton} events due to the observed differences in
flavor tagging efficiencies in the \jpsi\ sideband relative to \bflav\
and inclusive \jpsi\ Monte Carlo.  Conservatively, this error was
evaluated comparing the fit results with and without the correction.
The total \bcpkl\ specific systematics is evaluated by taking the
quadratic sum of the individual contributions (q).

\subsection{Summary of systematic uncertainties}

All individual systematic contributions described above and summarized
in Table \ref{tab:systglob} are added in quadrature. The dominant
source of systematic error in the measurement of \reZparflat\ is due
to our limited knowledge of the doubly-\ckm-suppressed decays, which
also contributes significantly to the other measurements. The limited
Monte Carlo statistics are a dominant source of systematics for
\absqop, \imZ\ and to a lesser extent to
\sgndGoverG. Residual charge asymmetries also dominate the systematics on \absqop.
Our limited knowledge of the beam spot and SVT alignment also reflects
significantly on \imZ\ and \sgndGoverG.  The systematic error on
\sgndGoverG\ also receives a non-negligible contribution from our
understanding of the resolution function. The systematic uncertainties
on \sgndGoverG\ and \absqop\ when \CPT\ is assumed to be a good symmetry
were evaluated similarly, and found to be, respectively, \mbox{$\pm
0.020$ and $\pm 0.012$.}

\section{\boldmath Summary and discussion of results}
\label{sec:summary}

The conventional analysis of mixing and \CP\ violation in the 
neutral \B\ meson system neglects possible contributions from several 
sources that are expected to be small.  These include the difference of
the lifetimes of the two neutral \B\ meson mass eigenstates, the
\CP- and \T-violating quantity $|q/p|-1$, which is proportional to
$\im (\Gamma_{12}/M_{12})$, and potential 
\CPT\ violation.  To measure or extract limits on these quantities
requires the full expressions for time dependence in mixing
and \CP\ violation and consideration of systematic issues that might
mimic fundamental asymmetries we seek to measure, like detector charge asymmetries, 
different resolution function for positive and negative \dt, and
doubly-\ckm-suppressed decays from both fully reconstructed final flavor states 
and non-leptonic tagging states.

Our analysis of approximately 31,000 fully reconstructed flavor
eigenstates and 2600 \CP\ eigenstates sets new limits on the difference
of decay widths of \Bz\ mesons and on the \CP, \T, and \CPT\ violation
intrinsic to $\BzBzb$ mixing. The six independent parameters
governing mixing \mbox{(\dM, \dGoverGd)}, \CPT/\CP\ violation \mbox{(\reZ, \imZ)} and
\CP/\T\ violation (\imlambcpbare, \absqop) are extracted from a single fit
of both fully reconstructed \CP\ and non-\CP events, tagged and
untagged.  This provides the sensitivity required to separate the
small effects we seek from asymmetries in detector response and from
potentially obscuring correlations in the decays of the two \B\ mesons. 
The preliminary results are

$$\begin{array}{
r@{\ \ =\ }r@{.}l@{\pm 0.}l@{{\rm (stat.)}\pm 0.}
l@{{\rm{(syst.)}}\ \ [}r@{,}l@{]}l}
\sgndGoverGd&-0&008&037&018&-0.084&0.068&~,\\
\absqop     &1 &029&013&011& 1.001&1.057&~,\\
\reZparflat &0 &014&035&034&-0.072&0.101&~,\\
\imZ        &0 &038&029&025&-0.028&0.104&~.
\end{array}$$
The values in square brackets indicate the 90\% confidence-level intervals. 
When estimating the limits we also evaluated multiplicative contributions to the systematic error,
adding in quadrature with the additive systematic uncertainties.
Assuming \CPT\ invariance the results are

$$\begin{array}{
r@{\ \ =\ }r@{.}l@{\pm 0.}l@{{\rm (stat.)}\pm 0.}
l@{{\rm{(syst.)}}\ \ [}r@{,}l@{]}l}
\sgndGoverGd&-0&009&037&020&-0.087&0.069&~,\\
\absqop     &1 &029&013&012& 1.000&1.058&~.
\end{array}$$
The parameters \dM\ and \imlambcpflat\ are allowed to float, so that recent \B-Factory \dM\ results
\cite{ref:dM-babar-had,ref:dM-babar-dstlnu,ref:dM-babar-dilep,ref:dM-belle}
and our $\sin2\beta$ analysis based on the same data sample
\cite{ref:sin2b-babar} provide a cross-check.
 The value of the \CP/\T-violating parameter
\imlambcpflat\ increases by $+0.012$ when \CPT\ violation is allowed in the fit. 

\begin{figure}[h!]
\begin{center}
\epsfig{file=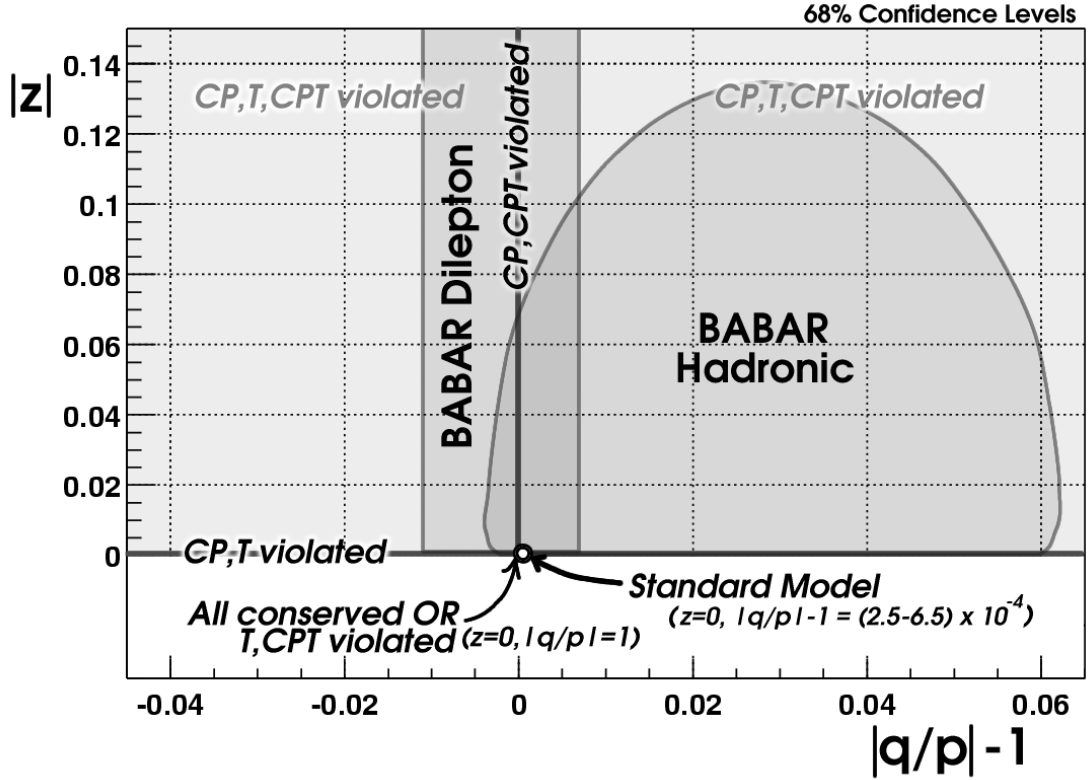,width=0.90\linewidth}
\end{center}
\caption{Favored regions at 68\% confidence level in the $(\absqop-1,|z|)$ plane
  determined by this analysis ("Hadronic") and by the \babar\ measurement of the
  dilepton asymmetry [20]. The axis labels reflect the requirements that both
  \CP\ and \T\ be violated if $\absqop\ne 1$ and that both \CP\ and \CPT\ be violated if
  $|z|\ne 0$. The region shown for this analysis is constrained to lie within
  the physical region $|z|\ge 0$ and is chosen to exclude the maximum range of
  alternative hypotheses given the {\em a priori} low probability of obtaining
  $|z|=0$ due to phase-space considerations. The dilepton measurement
  constrains $\absqop$ without assumptions on the value of $|z|$. The Standard
  Model expectation is obtained from Ref.~\cite{ref:absqopSM}.}\label{fig:zqp}
\end{figure}

The results are consistent with Standard Model expectations and with \CPT\ invariance.
To date, these are the best limits on the difference of decay widths
of \Bz\ mesons and the strongest test of \CPT\ invariance outside the
neutral kaon system \cite{ref:cptkaons}. The limit on \CP\ and \T\ violation 
in mixing is independent of and consistent with our previous
measurement based on the analysis of inclusive dilepton events
\cite{ref:babardileptonTviolation}. Fig.~\ref{fig:zqp} shows the 
results in the $(\absqop-1,|z|)$ plane, comparing them to the \babar\ measurement 
of $\absqop$ made with dileptons and to the Standard Model expectations.
All the other results are also
consistent with previous analyses
\cite{ref:dM-belle,ref:cleochid,ref:DELPHI,ref:otherCPTBtests}.  While
the Standard Model predictions for \dG\ and
$\absqop-1$ are still well below our current limits and no \CPT\ violation
is anticipated, higher precision measurements may still bring
surprises.

\section{\boldmath Acknowledgments}
\label{sec:acknowl}
We are grateful for the 
extraordinary contributions of our \pep2\ colleagues in
achieving the excellent luminosity and machine conditions
that have made this work possible.
The success of this project also relies critically on the 
expertise and dedication of the computing organizations that 
support \babar.
The collaborating institutions wish to thank 
SLAC for its support and the kind hospitality extended to them. 
This work is supported by the
US Department of Energy
and National Science Foundation, the
Natural Sciences and Engineering Research Council (Canada),
Institute of High Energy Physics (China), the
Commissariat \`a l'Energie Atomique and
Institut National de Physique Nucl\'eaire et de Physique des Particules
(France), the
Bundesministerium f\"ur Bildung und Forschung and
Deutsche Forschungsgemeinschaft
(Germany), the
Istituto Nazionale di Fisica Nucleare (Italy),
the Foundation for Fundamental Research on Matter (The Netherlands),
the Research Council of Norway, the
Ministry of Science and Technology of the Russian Federation, and the
Particle Physics and Astronomy Research Council (United Kingdom). 
Individuals have received support from 
the A. P. Sloan Foundation, 
the Research Corporation,
and the Alexander von Humboldt Foundation.

\appendix
\section{\boldmath Efficiency asymmetries}
\label{appendix:asymmetries}

The use of untagged data is essential to determining the asymmetries in
the tagging and reconstruction efficiencies.  To indicate how the various
samples enter we provide a simple example using only time-integrated quantities.
In practice we use a time-dependent analysis, which gives better precision because
it uses more information.
Suppressing the tag category $\alpha$, the signal or background
component, $j$, and writing the reconstruction efficiencies as
$r=r^j_{\Bz},\,{\overline r}=r^j_{\Bzb}$ and the
tagging efficiencies as
$t=t^{\alpha,j}_{\Bz},\,{\overline t}=t^{\alpha,j}_{\Bzb}$, Eq. (\ref{eq:numudef}) reads
\bea
 \nu & = & \frac{r-{\overline r}}{r+{\overline r}}\nn\\
 \mu & = & \frac{t-{\overline t}}{t+{\overline t}}~.
\label{eq:numudef2}
\eea
Using the numbers of signal events that are tagged and have a reconstructed \Bz ($X$), those
tagged and having a \Bzb ($Y$), those untagged with a reconstructed \Bz ($Z$) and finally
those untagged with a reconstructed \Bzb ($W$) we can determine the required asymmetries \cite{ref:babar-stwob-prd}.  
To see this, note that if the total number of $\BzBzb$ pairs is $N$, and neglecting \dG, \absqop\ and $z$ corrections,
there are $N_u=N(1+[1/(1+x_d^2)])/2$ unmixed
events (i.e. $\BzBzb$) and  $N_m=N(1-[1/(1+x_d^2)])/2$ mixed events (i.e. $\Bz\Bz$ or $\Bzb\Bzb$), 
where $x_d=\taub\dM$, so
\begin{eqnarray}
X&=&rt N_m/2 +r{\overline t}N_u/2\nonumber\\
Y&=&{\overline r}{\overline t} N_m/2 +{\overline r}t N_u/2\nonumber\\
Z&=&r(1-t) N_m/2 +r(1-\overline t)N_u/2\nonumber\\
W&=&{\overline r}(1-\overline t) N_m/2 +{\overline r}(1-t) N_u/2~.
\end{eqnarray}
Setting $U=X+Z$ and $V=Y+W$, we find
\begin{equation}
\nu=\frac{U-V}{U+V}\qquad ,~~~~~~ \mu=(1+x_d^2)\frac{(Y/V)-(X/U)}{(Y/V)+(X/U)}~.
\end{equation}
Corrections to these equations have to be applied due to non-zero values of \dG, $\absqop-1$ and $z$.
The use of untagged events is essential to the determination of $\nu$ and $\mu$. 

%
%
%
%

\newpage

\end{document}